\documentclass[useAMS,usenatbib]{mn2e}
\usepackage[pdftex]{graphicx} 
\usepackage[hidelinks]{hyperref}
\usepackage{xcolor} 
\usepackage{titlesec}
\usepackage{lscape}
\usepackage{calc}
\usepackage{amsmath}
\usepackage{amssymb}	
\usepackage{afterpage}
\usepackage{bm}

\titleformat{\paragraph}
{\it\normalsize}{\theparagraph}{1em}{}
\titlespacing*{\paragraph}
{10pt}{3.25ex plus 1ex minus .2ex}{1.5ex plus .2ex}

\hypersetup{colorlinks,linkcolor={red!50!black}, citecolor={blue!50!black},urlcolor={blue!80!black} }

\title[Gas Properties of the CGM and IGM]
{Properties of the ionized CGM and IGM: tests for galaxy formation models from the Sunyaev-Zel'dovich effect}

\author[S.H. Lim et al.] 
{S.H. Lim$^{1,2}$\thanks{E-mail:
slim@phas.ubc.ca}, 
David Barnes$^{3}$, 
Mark Vogelsberger$^{3}$, 
H.J. Mo$^{1}$, 
Dylan Nelson$^{4}$, 
\newauthor
Annalisa Pillepich$^{5}$, 
Klaus Dolag$^{6,7}$, and 
Federico Marinacci$^{8}$
\\ \\
$^{1}$Department of Astronomy, University of Massachusetts, Amherst MA 01003-9305, USA \\
$^{2}$Department of Physics and Astronomy, University of British Columbia, 6224 Agricultural Road, Vancouver, BC V6T 1Z1, Canada \\
$^{3}$Kavli Institute for Astrophysics and Space Research, Massachusetts Institute of Technology, Cambridge, MA 02139, USA \\
$^{4}$Universit\"{a}t Heidelberg, Zentrum f\"{u}r Astronomie, Institut f\"{u}r theoretische Astrophysik, Albert-Ueberle-Str. 2, 69120 Heidelberg, Germany \\
$^{5}$Max-Planck-Institut f\"{u}r Astronomie, K\"{o}nigstuhl 17, 69117 Heidelberg, Germany \\
$^{6}$Max-Planck-Institut f\"{u}r Astrophysik, Karl-Schwarzschild-Str. 1, D-85748 Garching, Germany \\
$^{7}$University Observatory Munich, Scheinerstra{\ss}e 1, 81679 Munich, Germany \\
$^{8}$Department of Physics and Astronomy, University of Bologna, Via Gobetti 93/2, I-40129 Bologna, Italy 
} 

\begin{document} 

\date{\today}

\pagerange{\pageref{firstpage}--\pageref{lastpage}}

\pubyear{2020}

\maketitle

\label{firstpage}

\begin{abstract} 
We present a comparison of the physical properties of the ionized 
gas in the circumgalactic (CGM) and intergalactic (IGM) media at $z\sim0$ between observations and 
four cosmological hydrodynamical simulations: Illustris, TNG300 of the IllustrisTNG 
project, EAGLE, and one of the Magneticum simulations. 
For the observational data, we use the gas properties that are inferred from 
cross-correlating the Sunyaev-Zel'dovich effect (SZE) from the {\it Planck} CMB maps with haloes and large-scale structure. Both 
the observational and simulation results indicate that 
the integrated gas pressure in haloes 
deviates from the self-similar case, showing that feedback impacts haloes with $M_{500}\sim 10^{12-13}\,{\rm M_\odot}$. 
The simulations predict that more than half the baryons are displaced from haloes, while the gas fraction inferred from our observational data roughly equals the cosmic baryon fraction throughout the $M_{500}\sim 10^{12-14.5}\,{\rm M_\odot}$ halo mass range. All simulations tested here predict that the mean gas temperature in haloes is about the virial temperature, while that inferred from the SZE is up to one order of magnitude lower than that from the simulations (and also from X-ray observations). 
While a remarkable agreement is found for the average properties of the IGM between the observation and some 
simulations, we show that their dependence on the large-scale tidal field can break the degeneracy between models that show similar predictions otherwise. Finally, we show that 
the gas pressure and the electron density profiles from simulations are not well described by a generalized NFW (GNFW) profile. Instead, we present a new model with a mass-dependent shape that fits the profiles accurately. 

\end{abstract} 

\begin{keywords} 
methods: statistical -- galaxies: formation -- galaxies: evolution -- galaxies: haloes.
\end{keywords}

\section[intro]{INTRODUCTION}
\label{sec_intro}

The current understanding of galaxy formation is significantly 
limited due to the complexity of the physical processes responsible 
for the interchange of mass and energy 
between galaxies and their surroundings \citep[e.g.][]{mo10}. 
Gas and dark matter fall into the potential wells of haloes, 
collapsed objects that form via gravitational instability from the initial matter 
fluctuations present in the early Universe. 
Gas is heated by the virial shock at the edge of dark matter haloes, but it 
eventually cools through radiation to form stars. 
Observations, however, have accumulated evidence for 
galactic-scale winds that can heat and return gas from 
galaxies to the circumgalactic medium (CGM; medium in the vicinity of dark matter haloes associated with 
galaxies) and intergalactic medium (IGM; diffuse medium between haloes), possibly driven 
by the stellar and AGN feedback 
\citep[e.g.,][]{steidel10, jones12, martin12, newman12, rubin14, 
heckman15, chisholm16, tumlinson17}. 
The gas returned to the surrounding medium is mixed 
with gas newly accreted on to haloes from the cosmic web. 
It is still an open question whether the wind energy is large 
enough to return the gas from haloes to the IGM, or whether 
a substantial portion of the gas recycles back on to galaxies 
\citep{stocke13, werk14, werk16, borthakur16, prochaska17, rudie19}. 
Clearly, a systematic investigation of the CGM and IGM 
properties will help to improve our understanding of feedback. 

With the advent of large CMB surveys, the Sunyaev-Zel'dovich 
effect \citep[SZE;][]{sunyaev72} 
provides a promising way to probe the CGM and 
IGM gas properties. The scattering of the CMB photons 
with the free electrons on their paths from  
the last scattering surface to the observer changes the CMB 
spectrum, which is called the Sunyaev-Zel'dovich effect. 
The SZE produced by the thermal motions of electrons is 
referred to as the thermal SZE (tSZE), while that produced 
by the bulk motions of electrons is called the kinetic 
SZE (kSZE). Recent studies have demonstrated that the tSZE and kSZE 
from observations can be used to characterize the properties of 
the ionized gas in the CGM and IGM \citep{hand12, pcxi, vanwaerbeke14, 
hojjati15, ma15, hill16, hill18, lim18a, lim18b, degraaff19, 
tanimura19, lim20}. Using the SZE to trace the gas 
has advantages that it can probe relatively low-density regions such as 
outskirt of haloes and the IGM compared to X-ray observations, and that 
the derived gas properties do not depend on the gas metallicity 
and ionization states unlike absorption line studies toward quasars. 

Cosmological hydrodynamical simulations offer a theoretical framework 
to study the physical processes involved in galaxy evolution, 
and the resulting properties of the CGM and IGM 
\citep[e.g.,][]{keres05, keres09, fg11a, somerville15, oppenheimer18}. 
Large simulations that trace the evolution of the matter in a box 
with a sidelength of hundreds of Mpc \citep[e.g.,][]{vogelsberger13, 
vogelsberger14a, crain15, schaye15, dolag16, mccarthy17, pillepich18b}, 
and high-resolution `zoom-in' simulations 
that focus on the evolution of individual haloes 
\citep[e.g.,][]{hopkins14, hopkins18, muratov15, fattahi16, sawala16, 
vdV16, aa17, grand17, hafen19}, incorporate the relevant 
physics to reproduce the properties of galaxies and gas from observations 
reasonably well \citep[e.g.,][]{fg10, fg11b, fg15, fg16, hummels13, nelson18b}. 
However, simulations that are known to reproduce a similar set of key observations 
have employed a wide range of physical models. 
In order to break the degeneracy between 
the models, we need to test the physical assumptions with more detailed observations. 
To this end, there have been studies that compared the properties 
of CGM and IGM between simulations and observations 
\citep{battaglia12a, biffi13, lebrun15, schaye15, barnes17, gupta17, mccarthy17, hill18, ayromlou20}.
However, comparisons based on X-ray observations have been, in most cases, limited to 
haloes with $M_{500}\geq 10^{13-13.5}\,{\rm M_\odot}$ 
\citep[e.g.,][]{biffi13, schaye15, barnes17, mccarthy17} while feedback is expected to strongly 
impact haloes with $M_{500}\leq 10^{13-13.5}\,{\rm M_\odot}$ because of their 
shallow potential well \citep[e.g.,][]{lebrun15, mccarthy17, ayromlou20}. A recent study by \citet{truong20} 
extended the comparisons to a lower mass but is restricted to a local volume out to $\sim 100\,{\rm Mpc}$. 
Comparisons of the SZE 
spanning a wide range of mass down to group-size haloes have been limited to 
a few observational studies \citep[e.g.][]{pcxi}, which 
are known to be sensitive to methods used to derive the properties from observational data 
\citep{lebrun15, hill18}.

In this paper, we compare the observational inference based on the SZE from \citet{lim18a, lim18b, lim20} 
with predictions from hydrodynamical simulations, providing comprehensive tests for galaxy formation models. 
The observational constraints we use have been obtained by employing a simultaneous matching of filtering 
to increase the signal-to-noise as well as to disentangle the projection 
effects of haloes along same line-of-sight (LOS), and found robust against systematic effects 
such as foreground/background fluctuations and beam smearing effect \citep{lim20}. 
We use the cross-correlations between the SZE signals,  
dark matter haloes and large-scale environments as tests. 
The mass scale probed in our analysis by haloes of $M_{500}=10^{12-14.5}\,{\rm M_\odot}$ 
is where AGN feedback leaves the most evident signatures, hence a fertile regime 
for testing feedback models (although some models predict that feedback 
impacts a lower mass regime; see e.g. \citealt{ayromlou20}). In fact, we find that the impact of 
AGN feedback on the SZE, including whether it only increases the gas temperature or blows the gas 
to the outskirt of haloes or even out of haloes, depends highly on mass as well as implemented models. 


This paper is organized as follows. We present the observational 
data and the simulations used for our analysis in Sec\,\ref{sec_data}. 
We present the comparisons between the observation and the simulations in Sec\,\ref{sec_results}. 
Finally, we summarize our findings and conclude in Sec\,\ref{sec_sum}. 
Throughout the paper, we assume the {\it Planck} cosmology \citep{pcxiii} with $\sigma_8=0.816,\, h=0.677,\, \Omega_m=0.309,$ and $\,\Omega_b=0.0486$. We scale accordingly the results from observation or simulations 
that assumed other cosmologies. We define haloes based on $R_{500}$, 
a radius within which the mean density is 500 times the critical density, and on $M_{500}$, 
the mass enclosed within $R_{500}$. We note that, in the present paper, we refer to the medium inside 
dark matter haloes as CGM while referring to the medium between haloes as IGM, and that our analysis 
is focused on the properties of their ionized component rather than their cold component.

\section[data]{DATA}
\label{sec_data}

\subsection{The Sunyaev-Zel'dovich effect}
\label{ssec_SZE}

As the CMB photons pass through groups or clusters, they interact with free electrons therein 
via inverse Compton scattering due to thermal motion of the electrons (thermal SZE or tSZE) or via Doppler 
effect due to bulk, kinetic motion of the electrons (kinetic SZE or kSZE), which results in a change in the 
energy spectrum of CMB. 

The change of the CMB spectrum caused by the tSZE is characterized through a dimensionless parameter by
\footnote{In this paper, we ignore the relativistic corrections to the SZE, which is about 8\% for the tSZE 
for the most massive end of the mass range probed in our analysis, and then decreases proportionally 
with decreasing mass \citep{remazeilles19, lee20}.}, 
\begin{eqnarray}\label{eq_tSZE}
\left(\frac{\Delta T}{T_{\rm CMB}}\right)_{\rm tSZE} = 
g(x) y \equiv 
g(x)\frac{\sigma_{\rm T}}{m_{\rm e}c^2} \int{P_{\rm e} {\rm d}l}, 
\end{eqnarray}
where $T_{\rm CMB}=2.7255\,{\rm K}$, $y$ is the Compton parameter, 
$g(x)=x \coth(x/2)-4$ is the conversion factor at a given 
$x\equiv h\nu/(k_{\rm B}T_{\rm CMB}$), $\sigma_{\rm T}$ is 
the Thompson cross-section, $m_{\rm e}$ is the electron rest-mass, 
$c$ is the speed of light, $P_{\rm e}=n_{\rm e}k_{\rm B}T_{\rm e}$ is the electron 
pressure
with $n_{\rm e}$ and $T_{\rm e}$ the number density and temperature, respectively, 
of the free electron, and finally ${\rm d}l$ is the path length along the 
given line-of-sight (LOS). 

Similarly, the temperature change in the CMB spectrum by the kSZE is given by 
\begin{eqnarray}\label{eq_kSZE}
\left(\frac{\Delta T}{T_{\rm CMB}}\right)_{\rm kSZE}
 = -\frac{\sigma_{\rm T}}{c} \int{n_{\rm e} (\bm{v}\cdot\hat{\bm r}) {\rm d}l}, 
\end{eqnarray}
where $\boldsymbol{v}$ is the velocity of the free electrons 
in bulk motion, and $\hat{\bm r}$ is the unit vector along a LOS.

\subsection{Observational constraints from the SZE}
\label{ssec_obs}

In this paper we use the observational constraints obtained by \citet{lim18a, lim18b, lim20}. 
They are a series of papers that studied the gas properties in the CGM and IGM in a coherent 
way using the {\it Planck} observation \citep{tauber10, pci} and a filtering approach to increase the signal-to-noise 
and properly account for the projection effects. Here we only highlight their main findings and parts of their methods 
necessary for the analysis presented in this paper, 
and refer the reader to the original papers for the details 
regarding their methods, validation tests, and comparisons with 
other observations. 

They used the group catalogs of 
\citet{yang05, yang07} and \citet{lim17}, which contains a total of about half a million 
groups from 2MRS, 6dF, SDSS, and 2dF covering nearly the entire sky, combined with 
the reconstructed velocity field \citep{wang12}, 
to extract the average SZE flux associated with haloes of mass from $10^{12}$ to $10^{15}\,{\rm M_\odot}$. 
They generated model maps that include observational effects and projection effects, and minimized the $\chi^2$ 
with respect to the {\it Planck} maps to constrain the models. 
To properly account for the beam smearing effect by the relatively large beam size of the {\it Planck} ranging 
between $5\arcmin$ and $31\arcmin$, they applied filters mimicking 
the {\it Planck} beam to their model maps. They assumed the universal 
pressure profile (UPP) of \citet{arnaud10} for tSZE, and a $\beta$-profile constrained from South Pole Telescope 
(SPT) clusters \citep{plagge10} for kSZE, to determine the amplitudes of SZE signals and to convert the results to 
relevant integrated flux. They have shown that their results are robust against various noise sources and 
systematic effects including residual background/foreground fluctuations, beaming, residual dust emission, 
and uncertainty in the velocity reconstruction, as well as properly account for the projection effects. 
Additionally, they demonstrated that the results are also robust against truncation of the filters 
at different radii, mass incompleteness of the halo catalog, moderate variations 
in the filter shape, and fluctuations of the background. They suspect that the results are 
insensitive to a moderate range of profiles because the beam size of the {\it Planck} 
does not resolve profiles for most of the targets. They found that the projection effects of 
larger haloes along LOSs, which are accounted for in their analysis, are significant for 
the signals from low-mass haloes. Using simulations, however, they also found that the method 
tends to over-estimate the kSZE flux for low-mass haloes by up to $20\%$, possibly because of 
the contamination by the projection of gas outside haloes along LOS. 
The projection effect is expected to depend on the baryonic 
processes such as the stellar and AGN feedback, with a higher projection 
effect from a stronger feedback. 

The impact of feedback is believed not to be confined within haloes, but also leaves its imprint 
in the IGM. For a better understanding of feedback, it thus is helpful to investigate the SZE 
signal of the IGM. For this reason, we also include the IGM properties constrained by \citet{lim18b} 
for our comparison. They adopted a similar approach as described above, i.e. they constructed 
and filtered model maps, and compared with the {\it Planck} maps to constrain their models by minimizing 
the $\chi^2$. They used the density field reconstructed on $1\,h^{-1}{\rm Mpc}$ scale 
for the Sloan Digital Sky Survey Data Release 7 \citep[SDSS DR7;][]{abazajian09} volume, by \citet{wang16}. 
The reconstruction uses haloes identified by the group finder for the SDSS 7 region as a tracer of the underlying 
density field. Specifically, the SDSS volume is partitioned into domains associated with each halo. Then the 
average profiles of haloes, calibrated from $N$-body simulations, are assumed to distribute the density field 
within each domain. Using simulation and mock catalogs, the method has been tested and found to be 
successful in reconstructing the density field. Thus no significant bias in the reconstructed density field 
for the observational data (with respect to the density field simply calculated by summing up the particles 
in each cube of a $1\,h^{-1}\,{\rm Mpc}$ side from simulations; see section\,\ref{ssec_P_rho}) is expected. 
\citet{lim18b} adopted and constrained a double power-law relation between the pressure and density of the IGM, 
\begin{eqnarray}\label{eq_pl}
P_{\rm e} = 
\begin{cases} 
A\times(\rho_{\rm m}/\rho_{\rm m,0})^{\alpha_1}, 
& \mbox{if } \rho_{\rm m} \leq \rho_{\rm m,0} \\ 
A\times(\rho_{\rm m}/\rho_{\rm m,0})^{\alpha_2}, 
& \mbox{if } \rho_{\rm m} > \rho_{\rm m,0}, 
\end{cases}
\end{eqnarray}
where $\rho_{\rm m,0}$ is a characteristic density at which a transition in the slope is set to occur, 
and have found the median parameter values of 
$\{\rho_{\rm m,0}/\,{\overline\rho}_{\rm m},\,\alpha_1,\,\alpha_2\}=\{3.0,\,1.7,\,2.2\}$. 
We refer the reader to L18b for more details of the posterior distribution of the parameters. 

They also investigated the dependence of the pressure - density relation on 
the large-scale 
tidal field. This is motivated by findings based on simulations that, 
among various quantities such as stellar mass, star formation rate, black 
hole mass, velocity dispersion, etc, the large-scale tidal field is the 
most crucial second parameter 
that affects the gas pressure \citep{lim18b}. Following \citet{wang11}, 
the halo-based tidal field for each grid-cell of $(1\,h^{-1}{\rm Mpc})^3$ is estimated as 
the halo tidal force exerted on the surface of a sphere along a direction $\boldsymbol{t}$, 
normalized by the self-gravity of the sphere, 
\begin{eqnarray}\label{eq_tidal}
f(\boldsymbol{t}) = \frac{\sum_{i}GM_iR_{\rm g} (1+3\cos 2\theta_i)/r_i^3}
{2GM_{\rm g}/R_{\rm g}^2}
\end{eqnarray}
where the summation is over all the haloes, 
$M_i$ is the mass of halo $i$, $R_{\rm g}=0.5 h^{-1}{\rm Mpc}$ 
is the radius of the sphere that approximates the grid-cell, 
$M_{\rm g}$ is the mass enclosed within 
the grid-cell in question, 
$r_i$ is the separation between the center of the grid-cell  
and the halo `$i$', and $\theta_i$ is the angle between 
$\boldsymbol{t}$ and $\boldsymbol{r}_i$. The tidal field satisfies  
$t_1+t_2+t_3=0$ where $t_1$, $t_2$, and $t_3$ ($t_1\geq t_2 \geq t_3$) 
are the eigenvalues of the tidal field tensor. \citet{wang11} 
showed that $t_1$ represents well 
the magnitude of the tidal field. We thus use $t_1$ 
to characterize the tidal field strength of the grid-cells. For the computation of tidal field strengths from 
the observation, \citet{lim18b} used haloes identified in the group catalog. As can be seen in equation \ref{eq_tidal}, 
$f(\boldsymbol{t})$ and $t_1$ can be computed in the same manner for observation as is for simulations 
once haloes are identified. To test the uncertainty introduced by the halo identification, we constructed 
a mock survey of SDSS and applied the same halo finder as used by \citet{lim17} and \cite{lim18b}. 
From the test, we found no meaningful change in the resulting tidal field strength compared to the case that we 
use the true haloes directly identified in simulations. Throughout the present paper, thus we use the observational 
data from \citet{lim18b} while for simulations we compute $t_1$ using the true haloes, ignoring 
the uncertainty of the halo finder.

\subsection{Hydrodynamical simulations} 
\label{ssec_sim}

For our analysis, we use a number of state-of-the-art cosmological gas simulations for galaxy evolution: Illustris, TNG300, EAGLE, and Magneticum (see below and Table~\ref{tab_sim} for details). These simulations 
adopt different numerical techniques, cosmological models, and 
different implementations of physical processes, to trace the evolution of the simulated
universe. All these simulations identify haloes using a friends-of-friends 
\citep[FoF;][]{huchra82, davis85} algorithm. 

\begin{table}
 \renewcommand{\arraystretch}{1.6} 
 \centering
 \begin{minipage}{71mm}
  \caption{List of the simulations used for comparisons. From left-to-right the columns show: simulation name, 
                comoving box size, baryonic particle mass, and dark matter particle mass. }
  \begin{tabular}{cccc}
\hline
Simulation & $L$ & $m_{\rm baryon}$ & $m_{\rm DM}$ \\
 & $[h^{-1}\,{\rm Mpc}]$ & $[{\rm M_\odot}]$ & $[{\rm M_\odot}]$ \\
\hline
\hline
Illustris & 75 & $1.6\times10^6$ & $6.3\times10^6$ \\  
TNG300 & 205 & $1.1\times10^7$ & $5.9\times10^7$ \\  
EAGLE & 67.8 & $1.8\times10^6$ & $9.7\times10^6$ \\  
Magneticum & 352 & $1.4\times10^8$ & $6.9\times10^8$ \\  
\hline
\\
\end{tabular}
\label{tab_sim}
\end{minipage}
\vspace{-5mm}
\end{table}

\subsubsection{Illustris}

The first simulation is Illustris \citep{vogelsberger14a, vogelsberger14b, 
genel14, sijacki15}, which 
was run with the moving-mesh code AREPO \citep{springel10}, assuming WMAP9 cosmology 
with $h=0.704,\, \Omega_m=0.273,$ and $\,\Omega_\Lambda=0.727$ \citep{hinshaw13}. 
The traced components include gas cells, dark matter particles, stars and 
stellar wind particles, and super-massive black holes. 
Sub-grid models are employed for the physical processes  
such as cooling \citep{katz96, wiersma09}, star formation 
\citep{springel03} with a \citet{chabrier03} initial mass function, 
stellar feedback \citep{vogelsberger13, torrey14}, and AGN 
feedback \citep{springeletal05, sijacki07}. 
For the detailed implementation, we refer the reader 
to \citet{vogelsberger13}. 
The free parameters in the models were constrained using 
a set of mostly $z=0$ observations of the galaxy populations. 
In this paper, we use Illustris-1, the fiducial 
run, that has a box size of $L=75\,h^{-1}\,{\rm Mpc}$ and contains 
$2\times (1820)^3$ initial gas and dark matter particles. 
The target baryon mass and dark matter particle mass are 
$m_{\rm baryon}= 1.6 \times 10^6 {\rm M_\odot}$ and 
$m_{\rm DM}= 6.3 \times 10^6 {\rm M_\odot}$, respectively. 
The gravitational softening length for the dark matter particles is $1.4\,{\rm kpc}$, for the gas cells is adaptive with a minimum at about $0.7\,{\rm ckpc}$.

\subsubsection{TNG300}

The IllustrisTNG project \citep[][]{marinacci18, naiman18, nelson18a, 
pillepich18b, springel18}, 
the successor of Illustris, is a series of hydrodynamical simulations run 
with the AREPO code including ideal magnetohydrodynamics \citep{pakmor11} and assuming 
the cosmological model given by \citet{pcxiii} with 
$\sigma_8=0.816,\, h=0.677,\, \Omega_m=0.309,$ and $\,\Omega_b=0.0486$. 
Here we use the TNG300-1 run, the largest volume of the IllustrisTNG project (TNG300 hereafter), which 
is sampled with 
$(2500)^3$ dark matter particles and $(2500)^3$ initial gas cells in a periodic box of 
$(205\,h^{-1}\,{\rm Mpc})^3$. 
The physical galaxy formation model of IllustrisTNG is an extension and improvement of 
the original Illustris model, and is detailed in the IllustrisTNG methods papers \citep{weinberger17, pillepich18a}. 
One of the major changes in IllustrisTNG is a new model of black-hole-driven 
kinetic feedback at low-accretion rates (referred to as a wind mode), compared to the original 
Illustris where thermal energy from the feedback is injected into surrounding gas in a form of `bubbles'. 
The target baryon mass and dark matter particle mass are 
$m_{\rm baryon}= 1.1 \times 10^7 {\rm M_\odot}$ and 
$m_{\rm DM}= 5.9 \times 10^7 {\rm M_\odot}$, respectively. 
The $z=0$ Plummer equivalent gravitational softening of the collisionless 
component, and the minimum comoving value of the adaptive gas gravitational 
softening are $1.5\,{\rm kpc}$ and $0.37\,{\rm ckpc}$, respectively.

\subsubsection{EAGLE}

The Evolution and Assembly of GaLaxies and their Environments 
\citep[EAGLE;][]{schaye15, crain15, mcalpine16}, 
run with a modified version of GADGET-3 
Smoothed Particle Hydrodynamics (SPH) code \citep{springel05}
tracks the evolution of gas, stars, dark matter, and massive black 
holes in a simulated universe, 
by implementing sub-grid models for cooling \citep{wiersma09}, 
star formation \citep{schaye08}, stellar and AGN feedbacks 
\citep{booth09, rosas-guevara16}. The 
models are parameterized and the model parameters are tuned to match 
observations including the 
stellar mass function and 
stellar mass-black hole mass relation at $z\sim0$. The simulation 
assumes the {\it Planck} cosmology \citep{pcxvi}. 
For this paper, we use the simulation run of 
the largest box, $(100\,{\rm Mpc})^3$, sampled by $2\times (1504)^3$ particles. 
The initial baryonic particle mass and dark matter particle mass are 
$1.8\times 10^6 {\rm M_\odot}$ and $9.7\times 10^6 {\rm M_\odot}$, respectively. 
The comoving Plummer-equivalent gravitational softening and the maximum 
physical softening length are roughly 
$2.7\,{\rm kpc}$ and $0.70\,{\rm ckpc}$, respectively.

\subsubsection{Magneticum}

The Magneticum simulations \citep[e.g.][]{dolag16} are a set of cosmological 
hydrodynamical simulations with various 
volumes and resolutions, performed with an improved version of GADGET-3. 
The simulations adopted a WMAP7 flat $\Lambda$CDM cosmology with 
$\sigma_8=0.809,\, h=0.704,\, \Omega_m=0.272,$ and $\,\Omega_b=0.0456$ 
\citep{komatsu11}. 
The simulations include a variety of physical processes such as cooling 
and star formation 
\citep{springel03}, black holes and 
AGN feedback \citep{fabjan10, hirschmann14}, 
and thermal conduction \citep{dolag04}. 
The results presented in this paper are produced from a particular 
run that has a box size of 
$L=352\,h^{-1}\,{\rm Mpc}$ and is sampled by $2\times (1584)^3$ particles. 
The dark matter particle mass and gas particle mass are 
$6.9\times 10^8 {\rm M_\odot}$ and $1.4\times 10^8 {\rm M_\odot}$, respectively. 
The softening lengths are $3.8\,{\rm kpc}$ commonly for dark matter and gas 
particles.


\section[results]{RESULTS}
\label{sec_results}

\subsection{The thermal energy of halo gas}
\label{ssec_Y500}

\begin{figure}
\includegraphics[width=1.02\linewidth]{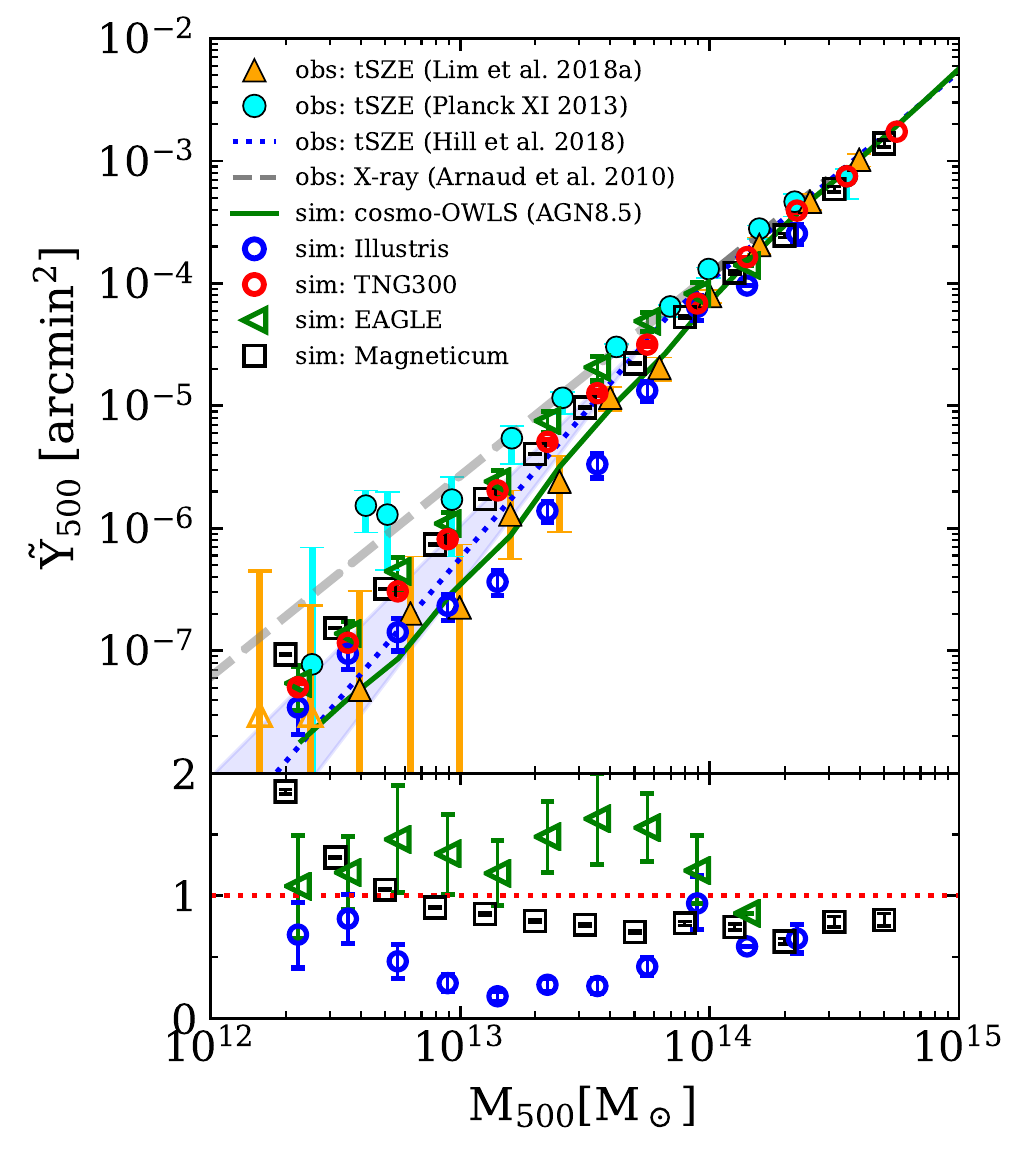}
\caption{Comparison of the tSZE flux from gas within 
$R_{500}$ of haloes, $\tilde{Y}_{500}$, between several observations 
(\citet[][up-pointing triangles]{lim18a}, \citet[][cyan dots]{pcxi}, 
\citet[][dotted]{hill18}, \citet[][dashed]{arnaud10} ), 
and simulations (cosmo-OWLS \citep[][green solid]{lebrun15}, 
Illustris (blue dots), TNG300 (red dots), EAGLE (left-pointing triangles), 
Magneticum (squares)). For the yellow triangles, the error bars 
represent the $68$ percentile ranges of the posterior distribution, 
and the unfilled symbols are used to indicate negative values. 
The error bars for the simulations indicate the $68$ percentile 
ranges of the mean obtained from $10,000$ bootstrap samples. In the lower panel, we 
show the ratio of the predictions from the simulations to that from TNG300. }
\label{fig_tSZE}
\end{figure}

Assuming a universal 
pressure profile (UPP) shape with amplitudes determined from $\chi^2$ minimization, \citet[][hereafter L18a]{lim18a} inferred from observational data the integrated tSZE flux within $R_{500}$, 
$Y_{500}$, defined by,  
\begin{eqnarray} \label{eq_Y500}
d_{\rm A}(z)^2 Y_{500} \equiv 
\frac{\sigma_{\rm T}}{m_{\rm e}c^2} \int_{R_{500}}{P_{\rm e}\,{\rm d}V}, 
\end{eqnarray}
where $d_{\rm A}(z)$ is the angular diameter distance to a halo at given redshift. 
At a fixed halo mass, $Y_{500}$ evolves with redshift as $E^{2/3}(z)$. Thus it is conventional 
to define a redshift-independent quantity scaled to $z=0$, 
\begin{eqnarray} \label{eq_Y500tilde}
\tilde{Y}_{500}\equiv Y_{500}E^{-2/3}(z)
\Big({d_{\rm A}(z)\over 500 {\rm Mpc}}\Big)^2, 
\end{eqnarray}
which is expected to be a function of only halo mass if the intrinsic tSZE flux follows a self-similar 
expectation across redshifts, i.e. the expectation from 
the virialization and a fixed baryon fraction at a given mass. 
The $\tilde{Y}_{500}$ thus inferred from the medians and $68$ percentile ranges 
of the posterior distribution of the model parameters from L18a are shown by the 
yellow triangles and the error bars, respectively, in Fig.\,\ref{fig_tSZE}.

The dashed line shows the self-similar case of 
\citet{arnaud10}, obtained from a combination of X-ray observations and simulations, 
that here we show extrapolated to a much lower mass than that probed in the original paper. The curve for the self-similar case in the extrapolated regime is 
shown in light gray while that in the mass scale directly probed in the original paper is in dark gray. 
We also present results from other observational studies, including 
\citet[][hereafter PCXI; cyan dots]{pcxi} and 
\citet[][blue dotted line]{hill18}. In order to account for 
different cosmologies assumed in the studies, we scale the results to the {\it Planck} 
cosmology with respect to the self similar case, 
$\int_{R_{500}}{P_{\rm e}\,{\rm d}V} \propto (\Omega_b/\Omega_m)\,h^{2/3}$. 
Using the {\it Planck} temperature maps, and locally brightest galaxies as a 
tracer of haloes, PCXI found that 
the tSZE flux follows the self-similar case, i.e. the thermal 
energy of hot gas relative to that of virialized haloes is independent of halo mass, 
whereas L18a reported a significantly lower thermal energy in lower-mass systems. 
L18a suspect that the discrepancy arises because PCXI did not fully take into 
account the projection 
effects of other haloes. PCXI tested both aperture photometry and matched 
filter to extract the flux, 
but in both cases they assumed flat local backgrounds to subtract. As demonstrated in 
\citet{vikram17}, however, the two-halo terms dominate the tSZE signals around haloes of 
$M_{200}\leq 10^{13-13.5}\,h^{-1}{\rm M_\odot}$, thus even a very small deviation 
from flat backgrounds can significantly change the estimation of the tSZE 
flux for those haloes. 
L18a confirmed that they recover the PCXI results when assuming flat 
backgrounds, implying that local background indeed changes with distance 
from halo centers due to the clustering of haloes. 
Taking into account the projection effects based on \citet{vikram17}, 
\citet{hill18} also found some evidence for a deviation of the relation 
from the self-similar case. As seen in 
Fig.\,\ref{fig_tSZE}, their results are consistent with the results from L18a even down to 
low-mass systems. Hill et al. used the {\it Planck} map and 
the halo catalog by \citet{yang07}, which is very 
similar to the data set that L18a used. 

Fig.\,\ref{fig_tSZE} compares the observational results with the predictions 
from the simulations described in the previous Section. To compute $\tilde{Y}_{500}$ from the simulations, we sum up 
the thermal energy of all ionized gas particles/cells that are associated with 
each simulated halo by the FoF algorithm and that are within a three-dimensional sphere of 
a radius $R_{500}$, without any temperature cut. The constants are multiplied 
according to equation \ref{eq_Y500} and \ref{eq_Y500tilde}. Note that $\tilde{Y}_{500}$ values 
from the simulations obtained this way do not contain any two-halo contribution or 
projection effects. 
The measurements from the simulations with different 
cosmologies are corrected to the {\it Planck} cosmology in the 
same way as was done for the observations. The error bars are obtained from 
$10,000$ bootstrap samples. It is clearly seen that all simulations considered in our analysis 
predict a certain degree of deviation from the self-similar case, 
as is seen in the case for the L18a observational constraints \citep[see also][]{battaglia12a}. 
The tSZE flux predicted by EAGLE and TNG300 are marginally consistent with that from L18a and  \citet{hill18}, 
within the uncertainties, which are typically up to a factor of $2$--$3$. 
Magneticum predicts higher tSZE flux for haloes with 
$M_{500} \leq 10^{12.5}\,{\rm M_\odot}$ than EAGLE and TNG300. 
On the other hand, Illustris, which implements a somewhat more violent AGN feedback than the other simulations 
\citep[see][]{genel15, pillepich18a, pillepich18b}, 
predicts a much lower electron pressure than the other simulations, and more 
than an order of magnitude lower pressure than the self-similar 
case for haloes with $M_{500}\sim 10^{13 -13.5}\,{\rm M_\odot}$. 
Those haloes are the 
systems that are believed to be strongly affected by the AGN feedback, suggesting that the discrepancies in the tSZE flux are due to the differing implementations of AGN feedback.

\begin{figure*}
\includegraphics[width=0.98\linewidth]{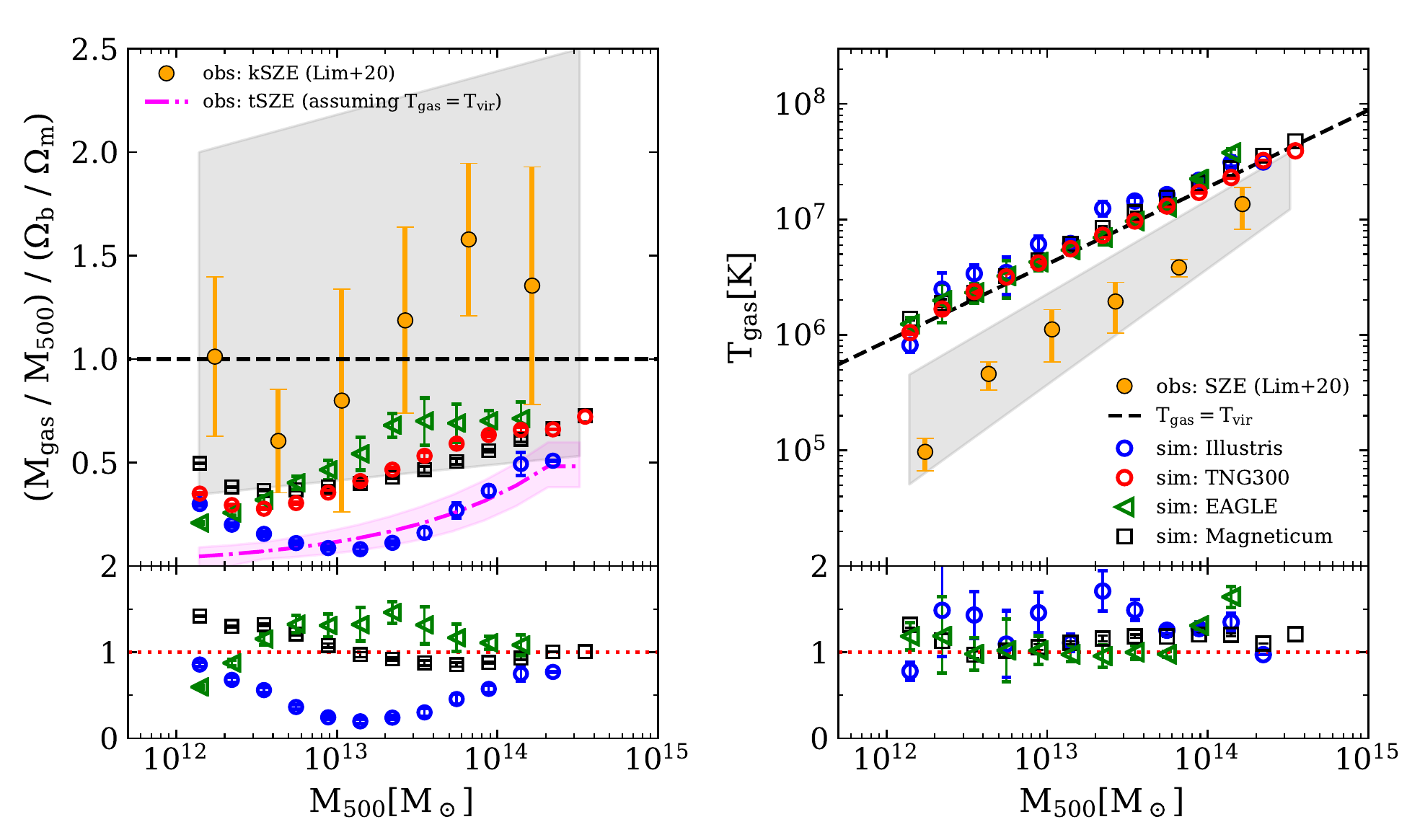}
\caption{Halo gas mass (left) and effective temperature (right) in comparison 
to the SZE-inferred observational constraints from \citet[][filled circle]{lim20}. 
The temperatures from the simulations are calculated as 
$T_{\rm eff}={\int_{R_{500}}{n_{\rm e}T_{\rm e}\,{\rm d}V}}\,/\,{\int_{R_{500}}{n_{\rm e}\,{\rm d}V}}$. 
The error bars for the observational result represent the dispersion of the 
estimate among the seven samples as described in \citet{lim20}. 
The shaded band spans the same dispersion among the seven samples but 
obtained with a power-law model from \citet{lim20}. 
The error bars for the simulations indicate the $68$ percentile 
ranges of the mean obtained from $10,000$ bootstrap samples. 
The dashed line shows the cosmic baryon fraction and the virial temperature 
$T_{\rm vir}=\mu m_{\rm p}GM_{500} / 2k_{\rm B}R_{500}$ 
in the left and right panels, respectively. 
In the left panel, the dot-dashed curve indicates the gas mass fraction 
inferred from the tSZE analysis by assuming ${\rm T_{gas}=T_{vir}}$, 
with the band showing the uncertainty. In the lower panels, we show the 
ratio of the predictions from the simulations to those from TNG300. } 
\label{fig_kSZE}
\end{figure*}

\subsection{The mass and temperature of halo gas}
\label{ssec_K500}

Using the amplitudes of the $\beta$-profile 
determined as described in section\,\ref{ssec_obs}, 
\citet[][hereafter L20]{lim20} inferred from observational data the total hot gas mass within $R_{500}$ as, 
\begin{eqnarray}
M_{\rm gas} = N_{\rm e, 500} \cdot \frac{2m_{\rm p}}{1+f_{\rm H}}, 
\end{eqnarray}
where $N_{\rm e, 500} =  \int_{R_{500}}{n_{\rm e}{\rm d}V}$, $f_{\rm H}=0.76$ 
is the hydrogen mass fraction, and $m_{\rm p}$ is the proton mass. 
Fig.\,\ref{fig_kSZE} shows the averages (the yellow circles) and the dispersions 
(the error bars) of the gas mass fraction from the seven observational samples described in L20. 
The shaded band spans the dispersion among the seven samples, inferred 
from a power-law model 
(see L20 for details). As seen in the figure, the inferred gas fraction 
is consistent with the cosmic baryon fraction (black dashed), thus there 
is no missing baryons in haloes. 
This is consistent with findings of \citet{hm15} 
that reported the detection of all baryons on halo scale 
from cross-correlation of SDSS galaxies with the kSZE of the {\it Planck} data. However, it is at odds with standard X-ray analyses, particularly for haloes below $\lesssim 10^{14}{\rm M_\odot}$ -- see below.

We compare the gas fraction from the kSZE with that inferred from the 
tSZE results by assuming that the gas is at the virial temperature, 
$T_{\rm vir}=\mu m_{\rm p}GM_{500} / 2k_{\rm B}R_{500}$ (where $\mu=0.59$ is the mean molecular weight), 
which is shown by the magenta dot-dashed line 
with the band showing the errors in the estimate. The inferred gas fraction 
from the tSZE is significantly lower particularly in low-mass systems than 
that from the kSZE. This implies 
that the effective temperature of the gas is much lower than the virial temperature as 
discussed in L20.  

We compare the observationally kSZE-inferred gas fractions with the predictions from the simulations. The latter is calculated in the simulations by summing up the ionized mass of 
all gas particles/cells within a three-dimensional sphere of $R_{500}$ that are associated with 
each simulated halo in question through the FoF, and dividing it by total halo mass and 
the cosmic baryon fraction. 
The error bars for the simulations are obtained from $10,000$ bootstrap samples. 
Overall, 
the simulations predict lower gas fractions than the observational results 
across the whole mass range considered here, up to by a factor of $\sim 4$ 
at the low-mass end of $M_{500}\sim 10^{12}{\rm M_\odot}$. 
This also means that the simulations predict $20-40\%$ 
lower baryons in low-mass systems relative to the cosmic fraction. 
Illustris has a much higher fraction of gas expelled out of the halo potential 
than the other simulations, possibly due to the stronger AGN feedback implemented \citep{genel14, pillepich18a}. 

\begin{figure}
\includegraphics[width=1.03\linewidth]{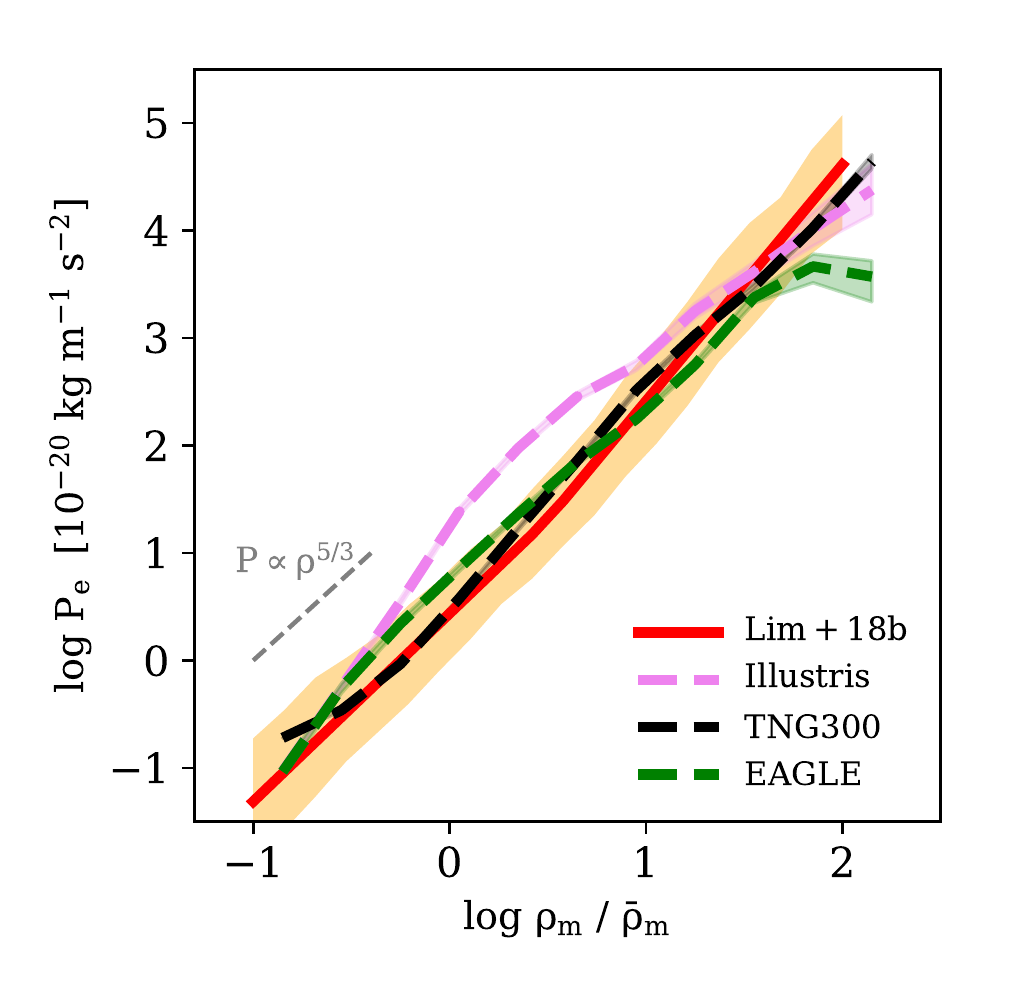}
\caption{The pressure - density relation of IGM gas, compared with 
that from the simulations. 
The red line shows the mean relation obtained with our method, with 
the orange band showing the $1\sigma$ dispersion estimated from the 
uncertainties in the constrained parameters. 
The violet, black, and green lines 
show the mean relation from Illustris, TNG300, 
and EAGLE, respectively, with the bands indicating the $68$ percentile 
ranges of the mean from $10,000$ bootstrap samples. }
\label{fig_P_rho}
\end{figure}

We also present the effective temperature, $T_{\rm eff}$, from L20 obtained by dividing the 
total tSZE flux with the total kSZE flux, in the right panel of Fig.\,\ref{fig_kSZE}. 
The inferred temperature (and 
also the inferred baryon fraction as they are connected to each other through 
the tSZE measurement) is in conflict with that reported from X-ray observations as pointed out 
and discussed in detail by L20. 
The temperature (baryon fraction) is found to be significantly lower (higher) than 
that from X-ray observations \citep[e.g.,][]{vikhlinin06, sun09, giodini09, gonzalez13, lovisari15}. 
As discussed in L20, this may be due to the difference between the mass-weighted temperature and X-ray 
temperature, the latter of which weights more the inner dense, hotter region. 
Simulations found that the X-ray temperature is indeed higher than 
the mass-weighted temperature \citep[e.g.][]{nagai07a, truong20}. The mass derived 
from X-ray also tends to be lower than the true halo mass by 
$0.2-0.3$\,dex on average 
\citep{nagai07a}, which results in a further bias to a higher temperature at 
a given mass. The net bias for the temperature comparison, however, 
is unclear since it depends on the gas distribution and 
properties which are not well understood in detail and largely model-dependent. 
The discrepancy, however, may be real and indicate that the gas responsible for the SZE and X-ray emission cannot 
be described well by a single component, but is composed of multiple components in different phases. 
\citet{wu20} found that both the SZE and X-ray data can be accommodated well simultaneously by a two-phase 
model where two gas components at different temperatures are assumed. They have shown that, in their model, 
the hot component is still found to be about at the virial temperature and comprises $20$--$60\%$ in mass, while 
the total (`hot'+`warm' components) gas mass fraction is close to the cosmic baryon fraction. The finding by 
\citet{wu20} thus indicates that the apparent discrepancy between the temperature inferred from the SZE and 
X-ray is not a conflict, but a reflection of multi-phases of the gas components, and arises from different weightings 
by the two observations of temperatures of the multiple components. 

Yet another uncertainty included in the comparison of the temperature is that of the group catalog used. 
While the group catalog adopted in L18a and L20 has been confirmed to yield unbiased halo mass estimate, 
it has a typical scatter of $0.2$--$0.3$\,dex in mass estimate. In a forthcoming paper (Lim et al. 2021 in prep.), 
the impact of the scatter in mass will be further studied in detail. 
Another source of uncertainty is the offset of the halo centers identified by the halo finder 
relative to the true center. The miscentering will result in an under-estimation 
of both the tSZE and kSZE signals. This will result in an under-estimation of 
the temperature because the tSZE profile is expected to decrease more rapidly 
as going away from the halo centers than the kSZE profile. Furthermore, due to the 
nature of the simultaneous matching, the results for different mass bins are 
correlated, with a slight under-(over-)estimation for massive haloes possibly 
leading to a significant over-(under-)estimation for lower-mass haloes 
(see L20). Finally, if the assumed profiles are not good approximations 
of true underlying profiles, that may introduce additional bias in the integrated fluxes.  
We investigate all these potential systematics in detail in a forthcoming paper 
(Lim et al. 2021 in prep.) where preliminary results show that all the systematics change the results 
within the $1\sigma$ uncertainty. 


For comparison, we estimate the effective temperature, 
$T_{\rm eff}=\frac{\int_{R_{500}}{n_{\rm e}T_{\rm e}\,{\rm d}V}}{\int_{R_{500}}{n_{\rm e}\,{\rm d}V}}$, 
from the simulations. The results are scaled with respect to the virial temperature, 
$T_{\rm vir}\propto h^{2/3}$, to correct for the different cosmologies adopted 
in the simulations. As seen in Fig.\,\ref{fig_kSZE}, right panel, 
all the simulations predict that the effective temperature of the gas is about 
the virial temperature, which is up to an order of magnitude higher than 
that inferred from the SZE analysis.  
It is interesting to note that the simulations with a wide range of feedback models 
predict very similar temperature of gas in haloes. Indeed, Fig.\,\ref{fig_kSZE} shows that 
the difference between the gas fractions predicted from the simulations is more significant than 
that in between the temperatures. In particular, EAGLE provides a simulation run for a smaller box 
(of a sidelength of 50\,{\rm Mpc}) with the `AGNdT9' model, where the AGN is set to heat the gas 
to a higher temperature of $10^9\,{\rm K}$, thus the feedback is more energetic and effective than 
in their fiducial model while the other parts of the simulation are kept the same. 
We directly confirm that their `AGNdT9' model predicts a much lower gas fraction at any given mass 
(typically by more than a factor of $2$) but predicts nearly the same temperature, relative to their fiducial 
model \citep[see also][]{barnes17}. All this indicates that the variations in the feedback models 
implemented in the simulations do not affect much the prediction about the temperature of halo gas, 
while they change the effectiveness at ejecting gas out of haloes significantly. 
This also indicates, in parallel, that the discrepancy between the simulation predictions and the SZE 
data shown here may be due to other factors in the numerical simulations than the models implemented 
for feedback, such as modelling multi-phase gas components.

\subsection{The pressure - density relation of IGM} 
\label{ssec_P_rho}

\begin{figure*}
\includegraphics[width=0.98\linewidth]{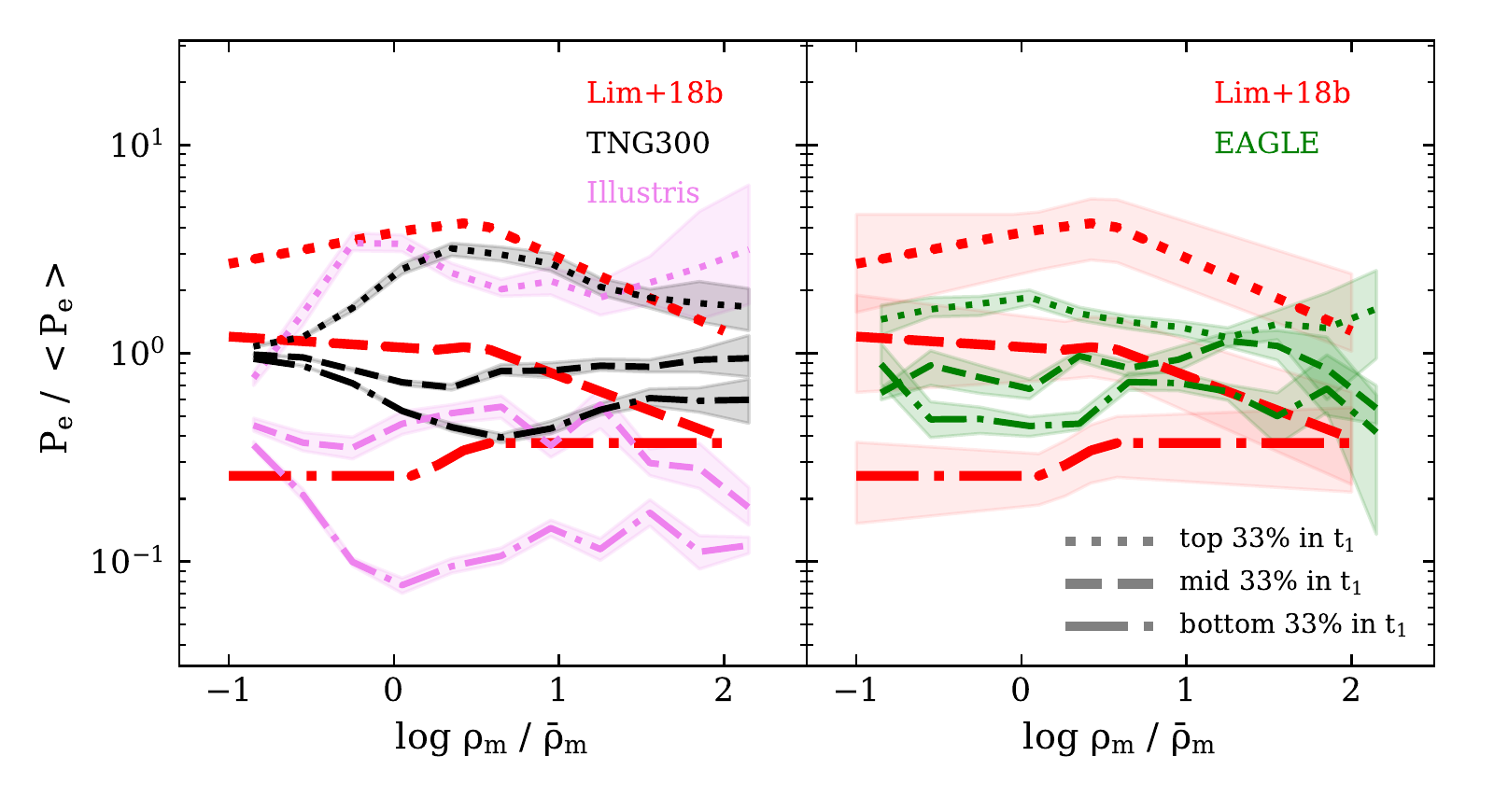}
\caption{The pressure - density relation obtained for the three
sub-samples of the grid-cells according to their ranking in the tidal 
field strength, $t_1$, at given density. 
The red lines show the mean relations obtained 
from the observational data, with the shaded bands indicating the $1\sigma$ 
dispersions estimated from the posterior distribution of the model parameters. 
The dispersions are presented only in the right panel, while they are same for 
the left panel, for visual clarity.  
The results are compared with the predictions from 
TNG300 and Illustris (left panel) and EAGLE (right panel). 
The shaded regions for the simulations indicate the $68$ percentile 
ranges of the mean obtained from $10,000$ bootstrap samples. 
For each case, the results are normalized by the mean relation from the 
whole sample of the grid-cells shown in Fig.\,\ref{fig_P_rho}.} 
\label{fig_tidal}
\end{figure*}

Fig.\,\ref{fig_P_rho} shows the pressure - density relation obtained by 
\citet[][hereafter L18b]{lim18b} with the power-law model as described in section\,\ref{ssec_obs}, 
by the red solid line with the band showing the $1\sigma$ scatter based on the posterior distribution. 
As one can see from the median values of the parameters, the relation closely 
follows that of an adiabatic equation of state, $P_{\rm e}\sim \rho^{5/3}$, but with a steeper 
slope in dense regions. 
The steeper slope may indicate an extra amount of heating sources available in dense regions due to 
feedback arising from star formation.  

In the figure, we also compare the observational result with the predictions from the simulations. The 
predictions are calculated by summing up the mass and thermal energy of particles 
within each grid-cell of $(1\,h^{-1}{\rm Mpc})^3$ from the simulations and dividing them 
by the volume of the grid-cells. Then the average over all grid-cells and its uncertainty are 
taken using $10,000$ bootstrap samples. As one 
can see, the agreement of the observational result with TNG300 and EAGLE is remarkable. 
Compared to TNG300, EAGLE predicts a higher thermal energy (up to a factor of 3) in the 
under-dense regions of $0.3 \leq \rho_{\rm m}/\,{\overline\rho}_{\rm m} \leq 1$, while it 
predicts a lower thermal energy in the regions of $10 \leq \rho_{\rm m}/\,{\overline\rho}_{\rm m}$, 
which are the regions corresponding to the cosmic structures. 
Illustris predicts a much higher thermal energy in the regions with 
$1 \leq \rho_{\rm m}/\,{\overline\rho}_{\rm m} \leq 10$, again, possibly due to 
the stronger radio-mode AGN feedback implemented in the simulation. 
One can convert the relations onto the temperature - density space, by assuming that 
each grid-cells has the average ionized gas mass fraction equal to the cosmic 
baryon fraction, as is confirmed to be the case from simulations (see L18b). 
The temperature thus estimated is lower than $10^4$K in the regions with 
$\rho_{\rm m} \leq \,{\overline\rho}_{\rm m}$, and increases with the density, 
up to $10^6$K in the regions with $\rho_{\rm m} \geq 100\,{\overline\rho}_{\rm m}$ 
which correspond to haloes. 

L18b also sub-sampled the grid-cells into three according to their tidal field strength, $t_1$, 
with each sub-sample containing a third of the total number of cells at a given density, 
and constrained the pressure - density relation for the three sub-samples jointly. 
We do the same calculation for the simulations for comparison by computing $t_1$ for 
the grid-cells in the simulations as described in section\,\ref{ssec_obs}, and by dividing 
the cells into three equal-sized sub-samples according to their ranking in $t_1$. 
The average relations between the pressure 
and matter density thus estimated from the observation and the simulations are 
shown in Fig.\,\ref{fig_tidal}. Both the simulations and the observation result 
show that the thermal energy of IGM at a given mass density is higher in the regions 
with the stronger tidal field. This may be due to that the regions with 
stronger tidal field are within or near massive structures, which cause the 
strong tidal field, where 
stronger stellar and AGN feedbacks are produced, resulting in more 
supply of heating. The tidal field dependence 
predicted by Illustris is much stronger than that from the observation 
as well as from the other simulations, which reflects the stronger AGN feedback 
implemented in Illustris. One can see that the prediction about the tidal field 
dependence distinguishes TNG300 and EAGLE with a much greater difference of up to 4--5$\sigma$ 
(particularly at $(1-10)\,{\overline\rho}_{\rm m}$), 
compared to the similarities shown mostly within 1--1.5$\sigma$ for the other predictions such as the gas contents 
of haloes and the total average pressure of IGM. TNG300 predicts much 
stronger dependence on the tidal field strength than EAGLE for the regions 
with $(1-10)\,{\overline\rho}_{\rm m}$, which correspond to cosmic 
filaments and sheets, and weaker dependence in under-dense regions. 
This may be due to that TNG300, unlike EAGLE, implemented two different models for 
two modes of the AGN feedback. Specifically, TNG300 employed a kinetic feedback model for 
the AGN at low-accretion rates while a thermal feedback model, as previously implemented in Illustris, 
was adopted for high-accretion rates. \citet{oppenheimer20} showed that, due to the two different 
modes of feedback, predictions for haloes at a given mass from TNG300 present highly bimodal distributions 
relative to EAGLE, which uses a single model for the AGN feedback. The much stronger dependence 
on the tidal field shown here also can be because the kinetic `wind-mode' feedback, which is far more 
efficient at ejecting halo gas than the thermal model \citep{weinberger17}, is more active in the region 
of a stronger tidal field. This gives an interesting hint that refining the balance between the modes and 
their strengths may improve the simulations toward more realistic models. Specifically, this could be explored by 
tuning the model parameters of feedback mode that is the main mechanism in the under-dense regions, to 
increase the feedback efficiency. 

Although simulations are consistent with the observation in that the 
thermal energy of gas is higher in the regions of the stronger tidal field, 
they fail to match the observation in detail. Compared to the observation, 
EAGLE predicts much weaker dependence on the tidal field in the density regimes 
corresponding to cosmic structures, while both TNG300 and EAGLE predict significantly weaker 
dependence in under-dense regions. All these clearly demonstrate the potential 
of leveraging cross-correlations with secondary parameters such as environment 
rather than just with haloes or galaxy systems, for providing stringent tests on 
galaxy models and thus for breaking degeneracies between them.

\subsection{The profiles of halo gas properties }
\label{ssec_profile}

\begin{figure*}
\includegraphics[width=1.\linewidth]{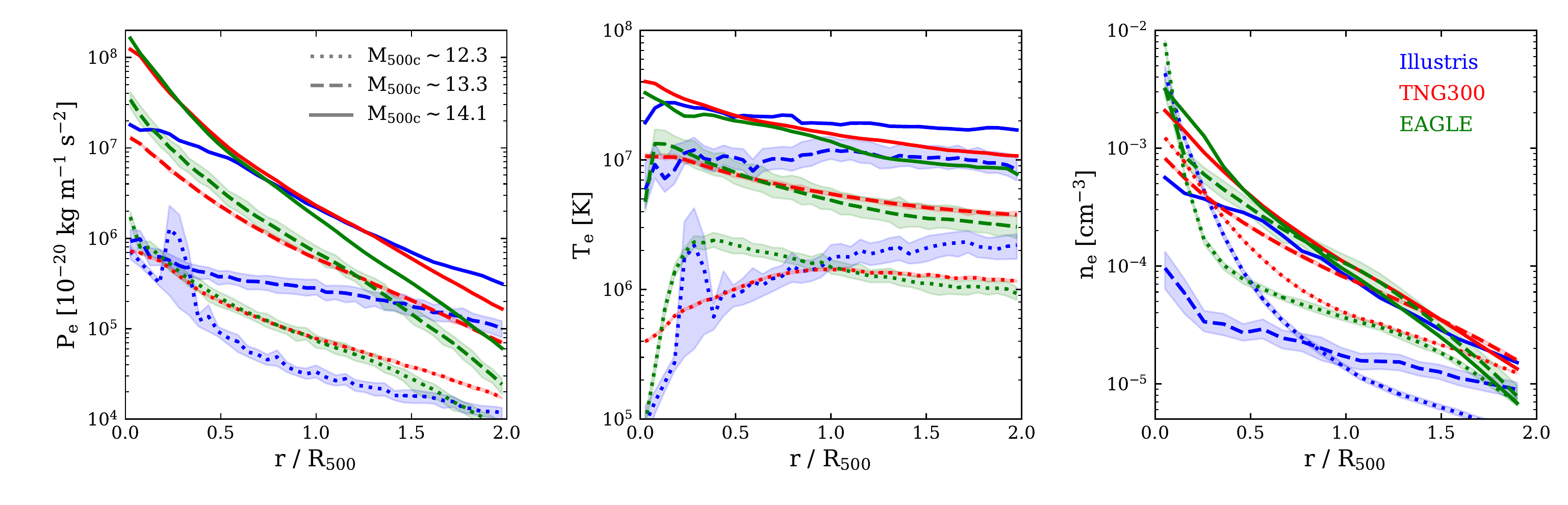}
\caption{Comparison of the pressure (left), 
effective temperature (middle), and electron number 
density (right) profiles of gas associated 
with haloes of 
different mass between the simulations. 
Different line styles were used for the profiles 
in haloes of different mass as indicated. 
The shaded regions for the simulations indicate the $68$ percentile 
ranges of the mean obtained from $10,000$ bootstrap samples. }
\label{fig_profiles}
\end{figure*}

\begin{figure*}
\includegraphics[width=1.02\linewidth]{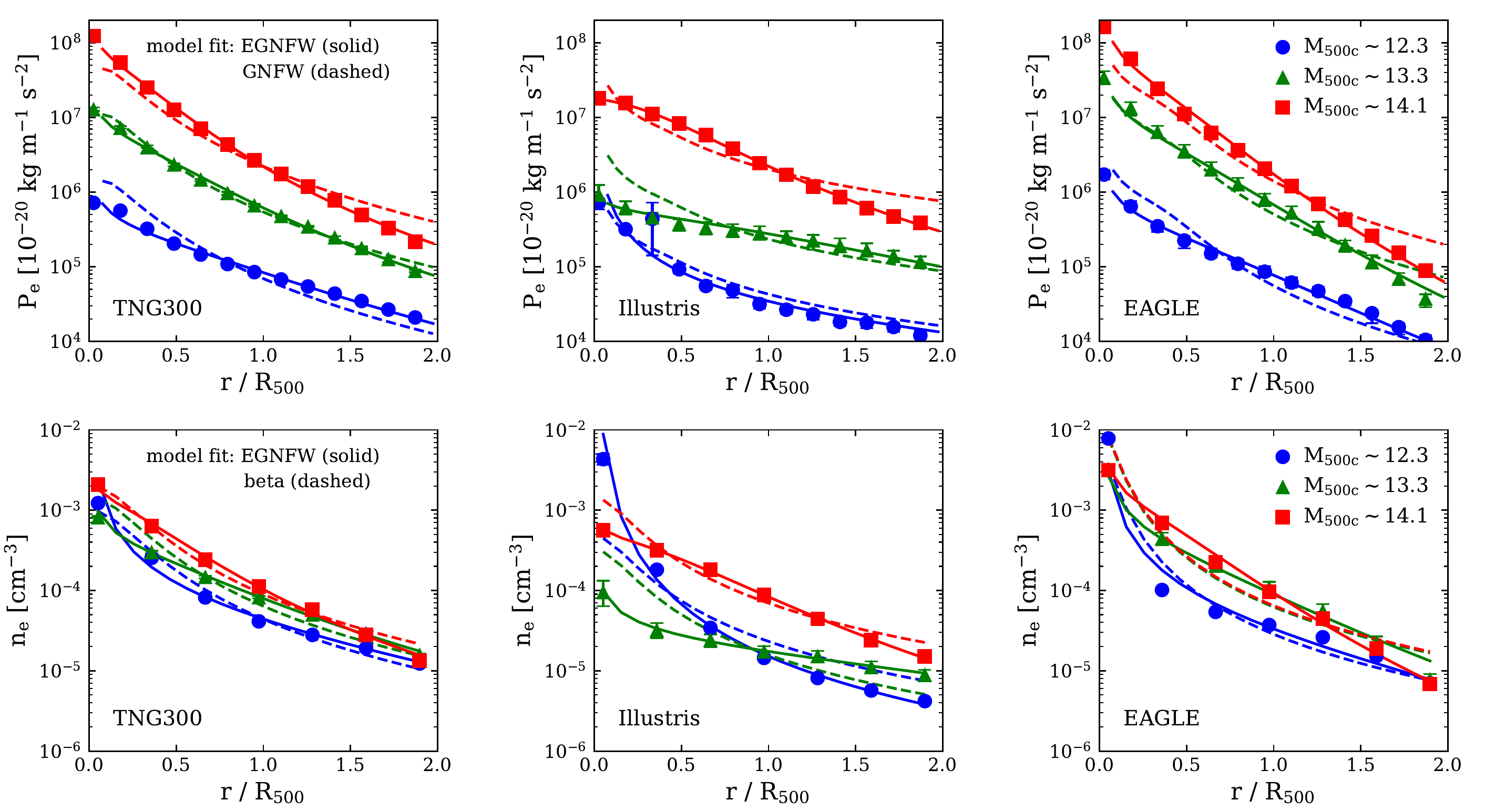}
\caption{Profiles of halo gas from the simulations (symbols) with model fits (lines). 
For clarity, the results are shown only for three halo mass ranges as indicated in the right panels. 
The error bars for the simulations indicate the $68$ percentile 
ranges of the mean obtained from $10,000$ bootstrap samples. 
({\it Upper}): The pressure profile fitted with the generalized NFW (GNFW; dashed lines) and 
extended GNFW (EGNFW; solid lines) models. 
({\it Lower}): The electron density profile fitted with the $\beta$-profile (dashed lines) and 
EGNFW (solid lines) model. } 
\label{fig_profile_modelfit}
\end{figure*}

We also investigate the profiles of gas properties including the pressure, 
temperature, and number density of free electrons, in proximity 
of galaxy haloes from the simulations. For the calculation of the profiles, 
we take into account all free electrons in each radial bin 
regardless of whether such electrons are associated to 
a halo in question or not. We find that accounting only for particles associated to haloes 
lowers the pressure and density profiles typically by less than $10\%$ ($50\%$) 
at $1.5R_{500}$ ($2R_{500}$). 
The pressure and density profiles 
are calculated by integrating the thermal energy and mass of free electrons, respectively, 
associated with gas particles/cells within each radial bins and by dividing them with the volumes 
enclosed in the corresponding shells. 
For the temperature profile, we calculate the effective temperature, 
$T_{\rm eff}(r)=\frac{\int_{\Delta r}{n_{\rm e}T_{\rm e}\,{\rm d}V}}{\int_{\Delta r}{n_{\rm e}\,{\rm d}V}}$, 
i.e. in the same way as can be obtained by combining the tSZE and kSZE from observation. 
The average profiles are shown in Fig.\,\ref{fig_profiles} for selected halo 
mass bins, with the bands showing the errors in the means obtained from 
$10,000$ bootstrap samples. The results are corrected to the {\it Planck} cosmology 
with respect to the virial temperature and universal baryon fraction. 

For massive haloes, Illustris predicts shallower profiles for the pressure and temperature 
compared to the other simulations, with less gas near the halo center, 
implying that the stronger AGN feedback implemented in 
Illustris removes the gas from halo center to the outskirts. 
That also explains the higher temperature 
in the outer region of haloes from Illustris. TNG300 also shows slightly shallower 
profiles than EAGLE, possibly due to the stronger `wind-mode' AGN feedback which pushes the gas 
to the outer region of haloes more effectively than the thermal feedback model implemented in EAGLE. 

For intermediate-mass haloes, the same trends as seen for the massive haloes continue 
but are much more strengthened. Illustris predicts more than an order of magnitude lower density 
in the inner region, and much less smooth temperature profile. For low-mass haloes, the 
difference among the simulations at $r\leq 0.5R_{500}$ are weaker than for intermediate-mass haloes, 
indicating that haloes strongly affected by different galaxy formation models are those with mass around 
$10^{13}\,{\rm M_\odot}$. It is interesting 
to note that, in Illustris, the gas in the outskirt of haloes is much less dense but is at 
a higher temperature than in TNG300 and EAGLE, implying that feedback implemented in Illustris expels gas completely 
out of low-mass haloes rather than just to outer regions that are still in proximity of the haloes. 
The temperature of the gas from Illustris is found to be much more stochastic compared to 
the other two simulations.

The functional form of the UPP is a generalized NFW (GNFW) model \citep{nagai07b} 
given by, 
\begin{eqnarray}\label{eq_GNFW}
\frac{P_{\rm e}(r)}{P_{\rm e,0}} = \frac{1}{(c_{500}\,r/R_{500})^\gamma [1+(c_{500}\,r/R_{500})^\alpha]^{(\beta-\gamma)/\alpha}}, 
\end{eqnarray}
where $P_{\rm e,0}$ is a normalization parameter, and 
the parameters $\gamma$, $\alpha$, and $\beta$ describe 
the slopes at $r\ll r_s$, at $r\sim r_s$, and at $r\gg r_s$, respectively, 
with $r_s=R_{500}/c_{500}$. It was shown that 
the pressure profile of X-ray groups and clusters from observations is well described 
by the GNFW model \citep[e.g.,][]{arnaud10, plagge10} 
but X-ray only allows to probe out to $r\sim R_{500}$ in most cases. 
Using a combination of X-ray observation and simulations, \citet{arnaud10} 
found the best parameter values of 
$\{\alpha,\,\beta,\,\gamma,\,c_{500}\}=\{1.05,\,5.49,\,0.308,\,1.18\}$. 

\begin{table*}
 \renewcommand{\arraystretch}{1.6} 
 \centering
 \begin{minipage}{110mm}
  \caption{The best-fit parameters of the EGNFW model for the pressure profiles, in comparison with other models from literature.}
  \begin{tabular}{ccccccc}
\hline
Profiles & $\alpha$ & $\beta$ & $\gamma_0$ & $\epsilon$ & $c_{500,0}$ & $\delta$ \\
\hline
\hline
Illustris & 2 (fixed) & 3.76 & 0.0545 & -0.803 & 1.20 & 0.517 \\  
TNG300 & 2 (fixed) & 4.01 & 0.456 & -0.0460 & 1.67 & 0.213 \\  
EAGLE & 2 (fixed) & 6.82 & 0.708 & 0.0338 & 1.02 & 0.129 \\  
\citet{arnaud10} & 1.05 & 5.49 & 0.308 & 0 (fixed) & 1.18 & 0 (fixed) \\  
\citet{lebrun15} & 2.02 & 3.84 & 1.08 & 0 (fixed) & 1.04 & 0.273 \\  
\hline
\\
\end{tabular}
\label{tab_profile_Pe}
\end{minipage}
\vspace{-5mm}
\end{table*}
\begin{table}
 \renewcommand{\arraystretch}{1.6} 
 \centering
 \begin{minipage}{77mm}
  \caption{The best-fit parameters of the EGNFW model for the density profile. }
  \begin{tabular}{cccccc}
\hline
Profiles & $\beta$ & $\gamma_0$ & $\epsilon$ & $c_{500,0}$ & $\delta$ \\
\hline
\hline
Illustris & 4.03 & 0.203 & -0.603 & 0.877 & 0.735 \\  
TNG300 & 3.50 & 0.288 & -0.379 & 1.52 & 0.306 \\  
EAGLE & 5.64 & 0.607 & -0.219 & 0.979 & 0.261 \\  
\hline
\\
\end{tabular}
\label{tab_profile_ne}
\end{minipage}
\vspace{-5mm}
\end{table}

\citet{lebrun15} (LB15 hereafter) found that the pressure distribution 
predicted from simulations is not well described by the GNFW for 
haloes of different mass, 
but requires an additional parameter for a mass dependence of $c_{500}$ instead, 
\begin{eqnarray}\label{eq_c500}
c_{500} = c_{500,0}\,(M_{500}/10^{14}{\rm M_\odot})^\delta\,. 
\end{eqnarray}
As seen in the equation, this extended GNFW (EGNFW, hereafter) suggests 
that the concentration parameter may be a function of mass, which is a 
reasonable expectation given potentially different impacts of feedback on haloes of 
different mass. LB15 indeed found a non-zero 
mass dependence with $\delta \sim 0.273$, indicating that the gas pressure is more concentrated 
in massive haloes than in lower-mass haloes. This may be because the gas at the center 
is blown out to the outskirts by galactic feedback more effectively in lower-mass haloes, 
resulting in a shallower profile. 

Here we find, however, that the profiles measured from TNG300, Illustris, and EAGLE are not described 
well even by the particular form of EGNFW used in LB15. This is because the asymptotic slope of 
the profiles in the innermost region, described by $\gamma$ in equation \ref{eq_GNFW}, 
is not universal for haloes of different mass, as clearly seen in Fig.\,\ref{fig_profiles}. 
This means that an additional parameter describing a mass-dependence in $\gamma$ is required. 

To this end, we suggest another form of the EGNFW model to describe the profiles: 
\begin{eqnarray}
P_{\rm e}(r), n_{\rm e}(r) \propto \frac{1}{(c_{500}\,r/R_{500})^\gamma [1+(c_{500}\,r/R_{500})^2]^{(\beta-\gamma)/2}}, 
\end{eqnarray}
where 
\begin{eqnarray}
\gamma &=& \gamma_0\,(M_{500}/10^{14}{\rm M_\odot})^\epsilon\,, 
\end{eqnarray}
with the concentration parameter following the same parameterization as in equation \ref{eq_c500}. 
Though the functional form is slightly different, this is conceptually very similar to the constrained fit 
dependent on mass in \citet{battaglia12b} and to the e-GNFW profile 
adopted in \citet{gupta17} to fit the pressure profiles from Magneticum, who found that the fit is 
greatly improved relative to fitting with the GNFW profile. Unlike the GNFW or LB15, we here choose 
to fix $\alpha$ because we find a significant degeneracy between $\alpha$ and $c_{500}$ when both 
parameters are free. This is as expected because, from equation \ref{eq_GNFW}, $\alpha$ 
determines the profile slope at $r\sim r_s$ while $c_{500}$ is directly related to $r_s$. The value of 
$\alpha=2$ was chosen to be very close to the value found by LB15 (see Table\,\ref{tab_profile_Pe}). 

We fit the pressure profiles measured from the simulations with 
the GNFW and our new EGNFW model, which are shown by the dashed and solid lines, respectively in 
the upper panels of Fig.\,\ref{fig_profile_modelfit}. The best-fit parameters 
for the EGNFW model are listed in Table\,\ref{tab_profile_Pe}. 
Unlike LB15, the normalization of the profile, $P_{\rm e,0}$, is treated 
as a free parameter in our model, the value of which is independently determined for haloes 
of different mass. We choose to do so because our main interest is the shape of profiles rather 
their amplitudes. 
It is clearly seen that the GNFW model, in which a profile shape is independent of mass, is 
insufficient to describe the pressure distribution accurately, because the pressure profiles in haloes 
of different mass clearly have different shapes thus are not matched only with normalizations. 
Specifically, it is seen, by comparing the dashed lines with the true profiles, that any mass-independent profiles 
including the GNFW would fail to match the profiles from the simulations because they over-estimate 
at $r>1.5\,R_{500}$ and at $r<0.5\,R_{500}$ for massive and low-mass haloes, respectively. 
The positive $\delta$'s from our EGNFW model fit indicate that the pressure profile is less concentrated in 
lower-mass haloes, consistent with the findings from previous studies such as LB15. This is because the gas is removed 
from the center to the outer radii by feedback more effectively in the lower-mass haloes due to their shallower potential well. 
Mass-independent profiles such as the GNFW, therefore, fail simply because they do not reflect the mass-dependent 
impact of feedback. 
The best-fit parameter values are remarkably similar to what \citet{gupta17} reported for Magneticum, 
although the mass range explored is significantly different. The model fit to the profile predicted by EAGLE 
is shown to have higher values for both $\beta$ and $\gamma_0$, which indicates steeper slopes both in the 
outer and inner regions of haloes, respectively, relative to the other two simulations. The profiles directly measured from 
the simulations indeed show that EAGLE predicts steeper slopes at both the inner and outer radii. 

We also fit the profile of electron number density with the $\beta$-profile and our EGNFW model. 
As seen in Fig.\,\ref{fig_profile_modelfit}, the shape of the profile strongly depends on mass, thus cannot be well 
described by a mass-independent profile such as the $\beta$-profile. As can be seen, instead, our new 
EGNFW model describes well the density profiles with the best parameters as listed in Table\,\ref{tab_profile_ne}. 
The negative $\epsilon$'s from our EGNFW model fit describe the steeper slope in the innermost region of lower-mass 
haloes, which cannot be described well by the $\beta$-profile or the EGNFW model from LB15. The concentration 
of the gas distribution is found to increase with increasing mass (i.e. positive $\delta$'s) as was the case for the 
pressure profiles. 




\section[summary]{SUMMARY AND DISCUSSION}
\label{sec_sum}

In this paper, we have compared the gas properties of the circumgalactic (CGM)  and intergalactic (IGM) media between an observational dataset and a selection of cosmological simulations, to 
explore the possibility of using the gas content within and around haloes to test models for physics 
of galaxy evolution. 

Specifically, we have used Sunyaev-Zel'dovich effect
(SZE) results obtained following \citet{lim18a, lim18b, lim20} as the observational data, 
and compared them with the predictions from four state-of-art cosmological hydrodynamical galaxy simulations at $z\sim 0$: Illustris, TNG300 of the IllustrisTNG 
project, EAGLE, and a Magneticum simulation run with a sidelength of $352\,h^{-1}\,{\rm Mpc}$. 
The observation data were obtained by cross-correlating {\it Planck} maps with haloes identified mainly in SDSS data by a halo finder, and with reconstructed large-scale environments such as density field, tidal field, and velocity field. 

For the tSZE signal from haloes (Fig.~\ref{fig_tSZE}), the observational results and the predictions from all the simulations considered in this analysis indicate a 
certain degree of deviation from the self-similar case, indicating that feedback 
impacts strongly the gas content of haloes with $M_{500}\sim 10^{12-13}\,{\rm M_\odot}$. This is in contrast with the conclusions by the \citet[][PCXI]{pcxi}, who find that the tSZE flux follows the self-similar case down to low-mass haloes with $M_{500}\sim 10^{12}\,{\rm M_\odot}$. We suspect that the reason for such difference lies in the assumption of flat local background adopted by PCXI \citep{lim18a}.  
Different simulations make different tSZE signal predictions because of the different implementation of the underlying physical model. Illustris, in particular, predicts a 
significantly lower thermal energy of gas in haloes with 
$M_{500}\sim 10^{13-13.5}\,{\rm M_\odot}$. This is believed to be due to 
the stronger AGN feedback adopted in that simulation, consistently with the findings and discussions of previous studies \citep{genel14, pillepich18a}. 
The predictions for $\tilde{Y}_{500}$ from TNG300 and EAGLE are remarkably similar, down to $M_{500}\sim 10^{12}\,{\rm M_\odot}$. 

The kSZE signal from the observational data (Fig.~\ref{fig_kSZE}, left) implies that the gas fraction in haloes is almost equal to the cosmic baryon fraction even in 
$M_{500}\lesssim 10^{13}\,{\rm M_\odot}$ haloes, in contrast with 
the predictions from simulations where the baryon fraction in those haloes is only $20-40\%$ of the cosmic fraction. There can be still, however, the residual 
contamination by gas along the LOSs between haloes not taken into account. 
Such contamination is found to be about $20\%$ 
according to simulations \citep{lim20} but is expected to depend too on the baryonic physics. 

Similarly, the effective temperature from the observations (Fig.~\ref{fig_kSZE}, right), obtained by dividing 
the tSZE flux with the kSZE flux, is also found to be up to one order of magnitude lower than that 
from the simulations. Notably, even the simulations with completely distinct feedback models produce very similar predictions, in that the halo gas is approximately at the virial temperature, whereas the prediction about the gas fraction varies much more significantly among the models. 
This indicates that the discrepancy between the predictions and the data may be due to 
other factors than the models implemented for feedback, such as modelling of multi-phase gas components. 

Moving to the gas that surrounds haloes, we have investigated the pressure-density relation of the IGM (Fig.~\ref{fig_P_rho}). The overall 
slope of the relation from both the observational data and the simulations closely follow that of an adiabatic equation of state. The observationally-inferred relation matches well the predictions 
from the simulations, except for Illustris that predicts a much higher thermal energy of IGM in the regions of intermediate density, this being too a manifestation of the strongly ejective AGN feedback at low-accretion rates adopted in Illustris. 

However, the dependence of the pressure-density relation on the tidal field strength (Fig.~\ref{fig_tidal}) is shown to have an important testing power for the models: all simulations fail to match the dependence for under-dense regions 
inferred from the observation in detail. On the other hand, the prediction by TNG300 matches the observation 
significantly better than EAGLE for the regions with an intermediate density of 
$(1-10)\,{\overline\rho}_{\rm m}$, which corresponds to cosmic structures 
such as filaments. We speculate that this may be because TNG300 employs two different modes to model AGN feedback, 
unlike the single-mode thermal model in EAGLE, and the `wind-mode' feedback -- that in TNG300 has been shown to be more effective than the thermal mode at displacing gas (e.g. \citealt{zinger20} and Pillepich et al. 2020) -- is more active 
in the regions with a stronger tidal field. Also, the mismatch of the TNG300 prediction with the observation 
for the under-dense region indicates an interesting hint that the current balance between the different 
modes of AGN feedback could potentially be adjusted to alleviate the tension. Specifically, the much stronger dependence on the tidal field strength 
found from the observation requires feedback mechanism(s) with clearly distinct strengths/modes in regions of different 
tidal fields in voids or under-dense part of the Universe. This might be achieved by fine-tuning the model parameters 
that control the feedback efficiency, i.e. breaking potentially remaining degeneracy in the parameter constraints 
by other observations. Although it is beyond the scope of this paper, this could be explored in future work with 
simulations where the observations presented in this paper are used to constrain the physical parameters jointly 
together with other observations previously used as the empirical constraints for simulations. 
All this shows that the tidal field dependence  of the IGM gas property can provide a significantly more 
stringent test to break the degeneracy between models that produce similar predictions otherwise. 

Finally, we have probed the profiles of gas properties in haloes predicted from the simulations (Figs.~\ref{fig_profiles},\ref{fig_profile_modelfit}). 
Different simulations predict differently shallow profiles of gas pressure and temperature, with more or less gas near the halo center, depending on the efficiency of the differently implemented AGN feedback models at displacing gas. 
The most prominent differences across models occur in intermediate-mass haloes with 
$M_{500}\sim 10^{13-13.5}\,{\rm M_\odot}$, which are hence a promising target to \textcolor{red}{test} AGN feedback. 

The profiles from the simulations are not well described by 
a generalized NFW (GNFW) profile or $\beta$-profile, both of which describe profiles with a universal shape regardless of mass unlike the profiles measured from the simulations. We demonstrate that a mass-dependent 
model is required to describe the simulated profiles accurately. Specifically, our new model (EGNFW), 
which incorporates such mass dependence in the innermost slope and 
concentration of the profiles, is shown to describe both the pressure and density profiles accurately out to $2R_{500}$. 

Our study clearly demonstrates the power of using the Sunyaev-Zel'dovich 
effect to test galaxy formation models. In particular, because most state-of-art simulations 
have their model parameters chosen to match a range of average properties 
of observed galaxies (in their stellar components, in particular), it is essential to 
take into account correlations in the properties of their gas, host haloes, and large-scale environment, 
in order to break the degeneracy among models. The haloes with mass between $10^{12}$ and 
$10^{14.5}$ studied in this paper are effectively impacted by AGN feedback and therefore offer a 
strong testing power for theory. With future CMB surveys such as the CMB-S4 
\citep{abazajian16, abitbol17, abazajian19}, 
Simons Observatory \citep{galitzki18a, galitzki18b, ade19}, and Toltec \citep{bryan18}, 
it is also expected that the SZE will allow us to probe the gas profiles 
of low-mass systems down to $M_{500}\sim 10^{12}\,{\rm M_\odot}$ 
by stacking (see also \citealt{battaglia17} and \citealt{battaglia19}). 
Similar comparisons as explored in this study can be applied to 
the future surveys to provide stringent tests on theoretical models of galaxy formation.

\section*{ACKNOWLEDGEMENTS}

We thank the referee for the constructive comments that improved the paper. 
We thank Huiyuan Wang and Xiaohu Yang for providing the group catalog and 
the reconstruction data. 
We thank the Planck collaboration for making the full-sky maps public, and the Virgo 
Consortium for making the EAGLE simulation data available. 
The EAGLE simulations were performed using the DiRAC-2 facility at Durham, 
managed by the ICC, and the PRACE facility Curie based in France at TGCC, 
CEA, Bruy\`eresle-Ch\^atel. 
We are also grateful to all the groups involved for making the Illustris
and IllustrisTNG project data available. 
SL thanks Eiichiro Komatsu and the Max Planck Institute for Astrophysics for 
hospitality, where this work was initiated. 
During this work, SL was supported in part by a CITA National Fellowship. 
FM acknowledges support through the program “Rita Levi Montalcini” of the Italian MIUR. 
KD acknowledges support by the Deutsche Forschungsgemeinschaft (DFG,
German Research Foundation) under Germany’s Excellence Strategy --
EXC-2094 -- 3907833. 

\section*{DATA AVAILABILITY}

The data underlying this article will be shared on reasonable request to the corresponding author.


\label{lastpage}


\begin{thebibliography}{}

\bibitem[\protect\citeauthoryear{Abazajian et al.}{2009}]{abazajian09} 
Abazajian K. N. et al., 2009, ApJS, 182, 543

\bibitem[\protect\citeauthoryear{Abazajian et al.}{2016}]{abazajian16} 
Abazajian K. N. et al., 2016, arXiv:1610.02743

\bibitem[\protect\citeauthoryear{Abazajian et al.}{2019}]{abazajian19} 
Abazajian K. N. et al., 2019, arXiv:1907.04473

\bibitem[\protect\citeauthoryear{Abitbol et al.}{2017}]{abitbol17} 
Abitbol M. H. et al., 2017, arXiv:1706.02464

\bibitem[\protect\citeauthoryear{Ade et al.}{2019}]{ade19} 
Ade P. et al., 2019, JCAP, 02, 056

\bibitem[\protect\citeauthoryear{Angl{\'e}s-Alc{\'a}zar et al.}{2017}]{aa17} 
Angl{\'e}s-Alc{\'a}zar D., Faucher-Gigu{\`e}re C.-A., Kere{\u s} D., 
Hopkins P. F., Quataert E., Murray N., 2017, MNRAS, 470, 4698

\bibitem[\protect\citeauthoryear{Arnaud et al.}{2010}]{arnaud10} 
Arnaud M., Pratt G. W., Piffaretti R., 
B\"ohringer H., Croston, J. H., Pointecouteau, E., 2010, A\&A, 517, 92

\bibitem[\protect\citeauthoryear{Ayromlou et al.}{2020}]{ayromlou20} 
Ayromlou M., Nelson D., Yates R.~M., Kauffmann G., Renneby M., White S.~D.~M., 2020, arXiv:2004.14390

\bibitem[\protect\citeauthoryear{Battaglia et al.}{2012a}]{battaglia12a} 
Battaglia N., Bond J.~R., Pfrommer C., Sievers J.~L., 2012, ApJ, 758, 74

\bibitem[\protect\citeauthoryear{Battaglia et al.}{2012b}]{battaglia12b} 
Battaglia N., Bond J.~R., Pfrommer C., Sievers J.~L., 2012, ApJ, 758, 75

\bibitem[\protect\citeauthoryear{Battaglia et al.}{2017}]{battaglia17} 
Battaglia N., Ferraro S., Schaan E., Spergel D.~N., 2017, JCAP, 2017, 040

\bibitem[\protect\citeauthoryear{Battaglia et al.}{2019}]{battaglia19} 
Battaglia N. et al., 2019, BAAS, 51, 297

\bibitem[\protect\citeauthoryear{Barnes et al.}{2017}]{barnes17} 
Barnes D. J. et al., 2017, MNRAS, 471, 1088

\bibitem[\protect\citeauthoryear{Biffi et al.}{2013}]{biffi13} 
Biffi V., Dolag K., B{\"o}hringer H., 2013, MNRAS, 428, 1395

\bibitem[\protect\citeauthoryear{Booth \& Schaye}{2009}]{booth09} 
Booth C. M., Schaye J., 2009, MNRAS, 398, 53

\bibitem[\protect\citeauthoryear{Borthakur et al.}{2016}]{borthakur16} 
Borthakur J. K. et al., 2016, ApJ, 833, 259

\bibitem[\protect\citeauthoryear{Bryan et al.}{2018}]{bryan18} 
Bryan S. et al., 2018, vol. 10708 of Society of Photo-Optical Instrumentation 
Engineers (SPIE) Conference Series, id. 107080J

\bibitem[\protect\citeauthoryear{Chabrier}{2003}]{chabrier03} 
Chabrier G., 2003, PASP, 115, 763

\bibitem[\protect\citeauthoryear{Chisholm et al.}{2016}]{chisholm16} 
Chisholm J., Tremonti C. A., Leitherer C., Chen Y., Wofford A., 2016,
MNRAS, 457, 3133

\bibitem[\protect\citeauthoryear{Crain et al.}{2015}]{crain15} 
Crain R. A. et al., 2015, MNRAS, 450, 1937

\bibitem[\protect\citeauthoryear{Davis et al.}{1985}]{davis85} 
Davis M., Efstathiou G., Frenk C. S., White S. D. M., 1985, ApJ, 292, 371

\bibitem[\protect\citeauthoryear{de Graaff et al.}{2019}]{degraaff19} 
de Graaff A., Cai Y.-C., Heymans C., Peacock J. A., 2019, A\&A, 624, 48

\bibitem[\protect\citeauthoryear{Dolag et al.}{2004}]{dolag04} 
Dolag K., Jubelgas M., Springel V., Borgani S., Rasia E., 2004, ApJ, 606, 97

\bibitem[\protect\citeauthoryear{Dolag et al.}{2016}]{dolag16} 
Dolag K., Komatsu E., Sunyaev R., 2016, MNRAS, 463, 1797

\bibitem[\protect\citeauthoryear{Fabjan et al.}{2010}]{fabjan10} 
Fabjan D., Borgani S., Tornatore L., Saro A., Murante G., Dolag K., 
2010, MNRAS, 401, 1670

\bibitem[\protect\citeauthoryear{Fattahi et al.}{2016}]{fattahi16} 
Fattahi A., et al., 2016, MNRAS, 457, 844

\bibitem[\protect\citeauthoryear{Faucher-Gigu{\`e}re et al.}{2010}]{fg10} 
Faucher-Gigu{\`e}re C.-A., Kere{\u s} D., Dijkstra M., 
Hernquist L., Zaldarriaga M., 2010, ApJ, 725, 633

\bibitem[\protect\citeauthoryear{Faucher-Gigu{\`e}re et al.}{2011a}]{fg11a} 
Faucher-Gigu{\`e}re C.-A., Kere{\u s} D., Ma C. P., 2011a, MNRAS, 417, 2982

\bibitem[\protect\citeauthoryear{Faucher-Gigu{\`e}re \& Kere{\u s}}{2011b}]{fg11b} 
Faucher-Gigu{\`e}re C.-A., Kere{\u s} D., 2011b, MNRAS, 412, 118

\bibitem[\protect\citeauthoryear{Faucher-Gigu{\`e}re et al.}{2015}]{fg15} 
Faucher-Gigu{\`e}re C.-A., Hopkins P. F., Kere{\u s} D., Muratov A. L.,
Quataert E., Murray N., 2015, MNRAS, 449, 987

\bibitem[\protect\citeauthoryear{Faucher-Gigu{\`e}re et al.}{2016}]{fg16} 
Faucher-Gigu{\`e}re C.-A., Feldmann R., Quataert E., Kere{\u s} D., 
Hopkins P. F., Murray N., 2016, MNRAS, 461, L32

\bibitem[\protect\citeauthoryear{Galitzki}{2018a}]{galitzki18a} 
Galitzki N., 2018a, arXiv:1810.02465

\bibitem[\protect\citeauthoryear{Galitzki et al.}{2018b}]{galitzki18b} 
Galitzki N. et al., 2018b, vol. 10708 of {\it Society of Photo-Optical Instrumentation 
Engineers (SPIE) Conference Series}, id. 1070804

\bibitem[\protect\citeauthoryear{Genel et al.}{2014}]{genel14} 
Genel S. et al., 2014, MNRAS, 445, 175

\bibitem[\protect\citeauthoryear{Genel et al.}{2015}]{genel15} 
Genel S., Fall S. M., Hernquist L., Vogelsberger M., Snyder G. F., 
Rodriguez-Gomez V., Sijacki D., Springel V., 2015, ApJ, 804, 40

\bibitem[\protect\citeauthoryear{Giodini et al.}{2009}]{giodini09} 
Giodini S. et al., 2009, ApJ, 703, 982

\bibitem[\protect\citeauthoryear{Gonzalez et al.}{2013}]{gonzalez13} 
Gonzalez A.~H., Sivanandam S., Zabludoff A.~I., Zaritsky D., 2013, ApJ, 778, 14

\bibitem[\protect\citeauthoryear{Grand et al.}{2017}]{grand17} 
Grand R. J. J. et al., 2017, MNRAS, 467, 179

\bibitem[\protect\citeauthoryear{Gupta et al.}{2017}]{gupta17} 
Gupta N., Saro A., Mohr J.~J., Dolag K., Liu J., 2017, MNRAS, 469, 3069

\bibitem[\protect\citeauthoryear{Hafen et al.}{2019}]{hafen19} 
Hafen Z. et al., 2019, MNRAS, 488, 1248

\bibitem[\protect\citeauthoryear{Hand et al.}{2012}]{hand12} 
Hand N. et al., 2012, Physical Review Letters, 109, 1101

\bibitem[\protect\citeauthoryear{Heckman et al.}{2015}]{heckman15} 
Heckman T. M., Alexandroff R. M., Borthakur S., Overzier R., Leitherer
C., 2015, ApJ, 809, 147

\bibitem[\protect\citeauthoryear{Hern\'andez-Monteagudo et al.}{2015}]{hm15} 
Hern\'andez-Monteagudo C., Ma Y.-Z., Kitaura F. S., Wang W., 
G\'enova-Santos R., Mac\'ias-P\'erez J., Herranz D., 2015, 
Physical Review Letters, 115, 191301

\bibitem[\protect\citeauthoryear{Hill et al.}{2016}]{hill16} 
Hill J. C., Ferraro S., Battaglia N., 
Liu J., Spergel D. N., 2016, Physical Review Letters, 117, 051301

\bibitem[\protect\citeauthoryear{Hill et al.}{2018}]{hill18} 
Hill J. C., Baxter E. J., Lidz A., Greco J. P., Jain B., 2018, 
Physical Review D, 97, 083501

\bibitem[\protect\citeauthoryear{Hinshaw et al.}{2013}]{hinshaw13} 
Hinshaw G. et al., 2013, ApJS, 208, 19

\bibitem[\protect\citeauthoryear{Hirschmann et al.}{2014}]{hirschmann14} 
Hirschmann M., Dolag K., Saro A., Bachmann L., Borgani S., Burkert A., 
2014, MNRAS, 442, 2304

\bibitem[\protect\citeauthoryear{Hojjati et al.}{2015}]{hojjati15} 
Hojjati A., McCarthy I. G., Harnois-Deraps J., Ma Y.-Z., Van Waerbeke L.,
Hinshaw G., Le Brun A. M. C., 2015, JCAP, 10, 047

\bibitem[\protect\citeauthoryear{Hopkins et al.}{2014}]{hopkins14} 
Hopkins P. F., Kere{\u s} D., Onorbe J., Faucher-Gigu{\`e}re C.-A.,
Quataert E., Murray N., Bullock J. S., 2014, MNRAS, 445, 581

\bibitem[\protect\citeauthoryear{Hopkins et al.}{2018}]{hopkins18} 
Hopkins P. F. et al., 2018, MNRAS, 480, 800

\bibitem[\protect\citeauthoryear{Huchra \& Geller}{1982}]{huchra82} 
Huchra J. P., Geller M. J., 1982, ApJ, 257, 423

\bibitem[\protect\citeauthoryear{Hummels et al.}{2013}]{hummels13} 
Hummels C. B., Bryan G. L., Smith B. D., Turk M. J., 2013,
MNRAS, 430, 1548


\bibitem[\protect\citeauthoryear{Jones et al.}{2012}]{jones12} 
Jones T., Stark D. P., Ellis R. S., 2012, ApJ, 751, 51

\bibitem[\protect\citeauthoryear{Katz et al.}{1996}]{katz96} 
Katz N., Weinberg D. H., Hernquist L., 1996, ApJS, 105, 19.

\bibitem[\protect\citeauthoryear{Kere{\u s} et al.}{2005}]{keres05} 
Kere{\u s} D., Katz N., Weinberg D. H., Dav{\'e} R., 2005, MNRAS, 363, 2

\bibitem[\protect\citeauthoryear{Kere{\u s} et al.}{2009}]{keres09} 
Kere{\u s} D., Katz N., Fardal M., Dav{\'e} R., Weinberg D. H., 2009, 
MNRAS, 395, 160

\bibitem[\protect\citeauthoryear{Komatsu et al.}{2011}]{komatsu11} 
Komatsu E. et al., 2011, ApJS, 192, 18

\bibitem[\protect\citeauthoryear{Le Brun et al.}{2015}]{lebrun15} 
Le Brun A. M. C., McCarthy I. G., Melin J.-B., 2015, MNRAS, 451, 3868

\bibitem[\protect\citeauthoryear{Lee et al.}{2020}]{lee20} 
Lee E., Chluba J., Kay S.~T., Barnes D.~J., 2020, MNRAS, 493, 3274

\bibitem[\protect\citeauthoryear{Lim et al.}{2017}]{lim17} 
Lim S. H., Mo H. J., Lu Y., Wang H., Yang X., 2017, MNRAS, 470, 2982

\bibitem[\protect\citeauthoryear{Lim et al.}{2018a}]{lim18a} 
Lim S. H., Mo H. J., Li R., Liu Y., Ma Y.-Z., Wang H., Yang X., 
2018a, ApJ, 854, 181

\bibitem[\protect\citeauthoryear{Lim et al.}{2018b}]{lim18b} 
Lim S. H., Mo H. J., Wang H., Yang X., 2018b, ApJ, 480, 4017

\bibitem[\protect\citeauthoryear{Lim et al.}{2020}]{lim20} 
Lim S. H., Mo H. J., Wang H., Yang X., 2020, ApJ, 889, 48


\bibitem[\protect\citeauthoryear{Lovisari et al.}{2015}]{lovisari15} 
Lovisari L., Reiprich T. H., Schellenberger G., 2015, A\&A, 573, 118

\bibitem[\protect\citeauthoryear{Ma et al.}{2015}]{ma15} 
Ma Y.-Z., Van Waerbeke L., Hinshaw G., Hojjati A., Scott D., 
Zuntz J., 2015, JCAP, 9, 046 

\bibitem[\protect\citeauthoryear{Marinacci et al.}{2018}]{marinacci18}
Marinacci F. et al., 2018, MNRAS, 480, 5113

\bibitem[\protect\citeauthoryear{Martin et al.}{2012}]{martin12} 
Martin C. L. et al., 2012, ApJ, 760, 127

\bibitem[\protect\citeauthoryear{McAlpine et al.}{2016}]{mcalpine16}
McAlpine S. et al., 2016, Astronomy and Computing, 15, 72

\bibitem[\protect\citeauthoryear{McCarthy et al.}{2017}]{mccarthy17} 
McCarthy I. G., Schaye J., Bird S., Le Brun A. M. C., 2017, 
MNRAS, 465, 2936

\bibitem[\protect\citeauthoryear{Mo et al.}{2010}]{mo10}
Mo H. J., van den Bosch F. C., White S. D. M., 2010, Galaxy Formation and 
Evolution. Cambridge University Press, New York, NY

\bibitem[\protect\citeauthoryear{Muratov et al.}{2015}]{muratov15}
Muratov A. L., Kere{\u s} D., Faucher-Gigu{\`e}re C.-A., Hopkins P. F.,
Quataert E., Murray N., 2015, MNRAS, 454, 2691

\bibitem[\protect\citeauthoryear{Nagai et al.}{2007a}]{nagai07a}
Nagai D., Vikhlinin A., Kravtsov A. V., 2007a, 655, 98

\bibitem[\protect\citeauthoryear{Nagai et al.}{2007b}]{nagai07b}
Nagai D., Kravtsov A. V., Vikhlinin A., 2007b, 668, 1

\bibitem[\protect\citeauthoryear{Naiman et al.}{2018}]{naiman18}
Naiman J. P. et al., 2018, MNRAS, 477, 1206

\bibitem[\protect\citeauthoryear{Nelson et al.}{2018a}]{nelson18a} 
Nelson D., et al., 2018, MNRAS, 475, 624

\bibitem[\protect\citeauthoryear{Nelson et al.}{2018b}]{nelson18b} 
Nelson D. et al., 2018, MNRAS, 477, 450

\bibitem[\protect\citeauthoryear{Newman et al.}{2012}]{newman12} 
Newman S. F. et al., 2012, ApJ, 761, 43

\bibitem[\protect\citeauthoryear{Oppenheimer}{2018}]{oppenheimer18} 
Oppenheimer B. D., 2018, MNRAS, 480, 2963

\bibitem[\protect\citeauthoryear{Oppenheimer et al.}{2020}]{oppenheimer20} 
Oppenheimer B. D. et al., 2020, ApJL, 893, L24

\bibitem[\protect\citeauthoryear{Pakmor et al.}{2011}]{pakmor11}
Pakmor R., Hachinger S., R{\"o}opke F. K., Hillebrandt W., 2011, A\&A, 528, 117

\bibitem[\protect\citeauthoryear{Pillepich et al.}{2018a}]{pillepich18a}
Pillepich A. et al., 2018a, MNRAS 473, 4077

\bibitem[\protect\citeauthoryear{Pillepich, et al.}{2018b}]{pillepich18b} 
Pillepich A. et al., 2018b, MNRAS, 475, 648

\bibitem[\protect\citeauthoryear{Plagge et al.}{2010}]{plagge10} 
Plagge T. et al., 2010, ApJ, 716, 1118

\bibitem[\protect\citeauthoryear{Planck Collaboration I}{2011}]{pci} 
Planck Collaboration I., 2011, A\&A, 536, 1

\bibitem[\protect\citeauthoryear{Planck Collaboration XI}{2013}]{pcxi} 
Planck Collaboration XI., 2013, A\&A, 557, 52

\bibitem[\protect\citeauthoryear{Planck Collaboration XVI}{2014}]{pcxvi} 
Planck Collaboration XVI., 2014, A\&A, 571, 16

\bibitem[\protect\citeauthoryear{Planck Collaboration XIII}{2016}]{pcxiii} 
Planck Collaboration XIII., 2016, A\&A, 594, 13

\bibitem[\protect\citeauthoryear{Prochaska et al.}{2017}]{prochaska17} 
Prochaska J. X. et al., 2017, ApJ, 837, 169

\bibitem[\protect\citeauthoryear{Remazeilles et al.}{2019}]{remazeilles19} 
Remazeilles M., Bolliet B., Rotti A., Chluba J., 2019, MNRAS, 483, 3459

\bibitem[\protect\citeauthoryear{Rosas-Guevara et al.}{2016}]{rosas-guevara16} 
Rosas-Guevara Y. M. et al., 2016, MNRAS, 462, 190

\bibitem[\protect\citeauthoryear{Rubin et al.}{2014}]{rubin14} 
Rubin K. H. R. et al., 2014, ApJ, 794, 156

\bibitem[\protect\citeauthoryear{Rudie et al.}{2019}]{rudie19} 
Rudie G. C., Steidel C. C., Pettini M., Trainor R. F., Strom A. L., 
Hummels C. B., Reddy N. A., Shapley A. E., 2019, ApJ, 885, 61

\bibitem[\protect\citeauthoryear{Sawala et al.}{2016}]{sawala16} 
Sawala T., et al., 2016, MNRAS, 457, 1931

\bibitem[\protect\citeauthoryear{Schaye \& Dalla Vecchia}{2008}]{schaye08} 
Schaye J., Dalla Vecchia C., 2008, MNRAS, 383, 1210

\bibitem[\protect\citeauthoryear{Schaye et al.}{2015}]{schaye15} 
Schaye J. et al., 2015, MNRAS, 446, 521

\bibitem[\protect\citeauthoryear{Sijacki et al.}{2007}]{sijacki07}
Sijacki D., Springel V., Di Matteo T., Hernquist L., 2007, MNRAS, 380, 877

\bibitem[\protect\citeauthoryear{Sijacki et al.}{2015}]{sijacki15}
Sijacki D., Vogelsberger M., Genel S., Springel V., Torrey P., 
Snyder G. F., Nelson D., Hernquist L., 2015, MNRAS, 452, 575

\bibitem[\protect\citeauthoryear{Somerville et al.}{2015}]{somerville15} 
Somerville R. S., Dav{\'e} R., 2015, ARAA, 53, 51

\bibitem[\protect\citeauthoryear{Springel \& Hernquist}{2003}]{springel03} 
Springel V., Hernquist L., 2003, MNRAS, 339, 289

\bibitem[\protect\citeauthoryear{Springel}{2005}]{springel05} 
Springel V., 2005, MNRAS, 364, 1105

\bibitem[\protect\citeauthoryear{Springel et al.}{2005}]{springeletal05} 
Springel V. et al., 2005, Nature, 435, 629

\bibitem[\protect\citeauthoryear{Springel}{2010}]{springel10} 
Springel V., 2010, MNRAS, 401, 791

\bibitem[\protect\citeauthoryear{Springel}{2018}]{springel18} 
Springel V. et al., 2018, MNRAS, 475, 676

\bibitem[\protect\citeauthoryear{Steidel et al.}{2010}]{steidel10} 
Steidel C. C., Erb D. K., Shapley A. E., Pettini M., Reddy N.,
Bogosavljevi{\'c} M., Rudie G. C., Rakic O., 2010, ApJ, 717, 289

\bibitem[\protect\citeauthoryear{Stocke et al.}{2013}]{stocke13} 
Stocke, J. T. et al., 2013, ApJ, 763, 148

\bibitem[\protect\citeauthoryear{Sun et al.}{2009}]{sun09} 
Sun M., Voit G.~M., Donahue M., Jones C., Forman W., Vikhlinin A., 2009, ApJ, 693, 1142

\bibitem[\protect\citeauthoryear{Sunyaev \& Zeldovich}{1972}]{sunyaev72} 
Sunyaev R. A., Zeldovich Y. B., 1972, 
Comments on Astrophysics and Space Physics, 4, 173

\bibitem[\protect\citeauthoryear{Tanimura et al.}{2019}]{tanimura19} 
Tanimura H., Hinshaw G., McCarthy I. G., Van Waerbeke L., 
Ma Y.-Z., Mead A., Hojjati A., Tr\"oster T., 2019, MNRAS, 483, 223

\bibitem[\protect\citeauthoryear{Tauber et al.}{2010}]{tauber10} 
Tauber J. A. et al., 2010, A\&A, 520, 1

\bibitem[\protect\citeauthoryear{Torrey et al.}{2014}]{torrey14} 
Torrey P., Vogelsberger M., Genel S., Sijacki D., Springel V., 
Hernquist L., 2014, MNRAS, 438, 1985

\bibitem[\protect\citeauthoryear{Truong et al.}{2020}]{truong20} 
Truong N. et al., 2020, MNRAS, 494, 549

\bibitem[\protect\citeauthoryear{Tumlinson et al.}{2017}]{tumlinson17} 
Tumlinson J., Peeples M. S., Werk J. K., 2017, ARAA, 55, 389

\bibitem[\protect\citeauthoryear{van de Voort et al.}{2016}]{vdV16} 
van de Voort F. et al., 2016, MNRAS, 463, 4533

\bibitem[\protect\citeauthoryear{Van Waerbeke et al.}{2014}]{vanwaerbeke14} 
Van Waerbeke L., Hinshaw G., Murray N., 2014, Physical Review D, 89, 023508

\bibitem[\protect\citeauthoryear{Vikhlinin et al.}{2006}]{vikhlinin06} 
Vikhlinin A., Kravtsov A., Forman W., Jones C., Markevitch M., Murray S.~S., Van Speybroeck L., 2006, ApJ, 640, 691

\bibitem[\protect\citeauthoryear{Vikram et al.}{2017}]{vikram17} 
Vikram V., Lidz A., Jain B., 2017, MNRAS, 467, 2315

\bibitem[\protect\citeauthoryear{Vogelsberger et al.}{2013}]{vogelsberger13} 
Vogelsberger M., Genel S., Sijacki D., Torrey P., Springel V., 
Hernquist L., 2013, MNRAS, 436, 3031

\bibitem[\protect\citeauthoryear{Vogelsberger et al.}{2014a}]{vogelsberger14a} 
Vogelsberger M. et al., 2014a, MNRAS, 444, 1518

\bibitem[\protect\citeauthoryear{Vogelsberger et al.}{2014b}]{vogelsberger14b} 
Vogelsberger M. et al., 2014b, Nature, 509, 177

\bibitem[\protect\citeauthoryear{Wang et al.}{2011}]{wang11} 
Wang H., Mo H. J., Jing Y. P., Yang X., Wang Y., 2011, MNRAS, 413, 1973

\bibitem[\protect\citeauthoryear{Wang et al.}{2012}]{wang12} 
Wang H., Mo H. J., Yang X., van den Bosch F. C., 2012, MNRAS, 420, 1809

\bibitem[\protect\citeauthoryear{Wang et al.}{2016}]{wang16} 
Wang H. et al., 2016, ApJ, 831, 164

\bibitem[\protect\citeauthoryear{Weinberger et al.}{2017}]{weinberger17} 
Weinberger R. et al., 2017, MNRAS, 465, 3291

\bibitem[\protect\citeauthoryear{Werk et al.}{2014}]{werk14} 
Werk J. K. et al., 2014, ApJ, 792, 8

\bibitem[\protect\citeauthoryear{Werk et al.}{2016}]{werk16} 
Werk J. K. et al., 2016, ApJ, 833, 54

\bibitem[\protect\citeauthoryear{Wiersma et al.}{2009}]{wiersma09} 
Wiersma R. P. C., Schaye J., Smith B. D., 2009a, 
MNRAS, 393, 99

\bibitem[\protect\citeauthoryear{Wu et al.}{2020}]{wu20} 
Wu X., Mo H., Li C., Lim S., 2020, ApJ, 903, 26

\bibitem[\protect\citeauthoryear{Yang et al.}{2005}]{yang05} 
Yang X., Mo H. J., van den Bosch F. C., Jing, Y. P., 2005, MNRAS, 356, 1293

\bibitem[\protect\citeauthoryear{Yang et al.}{2007}]{yang07} 
Yang X., Mo H. J., van den Bosch F. C., Pasquali A., Li C., Barden M., 
2007, ApJ, 671, 153

\bibitem[\protect\citeauthoryear{Zinger et al.}{2020}]{zinger20} 
Zinger E. et al., 2020, MNRAS, 499, 768
























\end{thebibliography}
\end{document}